%% file: AGN_catalog_arxiv.tex
\documentclass[twocolumn, times, tighten]{aastex62}
\usepackage{graphicx,amsmath,amssymb}
\usepackage{multirow} 
\usepackage{enumitem}

\newcommand{\bha}{H\ensuremath{\alpha ^{\rm b}}}

\newcommand{\civ}{C\,{\footnotesize IV}}
\newcommand{\chisq}{\ensuremath{\chi^2}}

\newcommand{\flux}{erg~s$^{-1}$~cm$^{-2}$}

\newcommand{\feii}{{\rm Fe\,{\footnotesize II}}}

\newcommand{\ha}{H\ensuremath{\alpha}}
\newcommand{\hb}{H\ensuremath{\beta}}
\newcommand{\hc}{H\ensuremath{\gamma}}

\newcommand{\hii}{H\,{\footnotesize II}}
\newcommand{\heii}{He\,{\footnotesize II} }

\newcommand{\kms}{\ensuremath{\mathrm{km~s^{-1}}}}

\newcommand{\lum}{erg~s\ensuremath{^{-1}}}
\newcommand{\lbol}{\ensuremath{L\mathrm{_{bol}}}}

\newcommand{\ledd}{\ensuremath{L\mathrm{_{Edd}}}}
\newcommand{\lratio}{\ensuremath{\lbol/\ledd}}

\newcommand{\lbha}{\ensuremath{L_{\rm \bha}}}

\newcommand{\lfive}{\ensuremath{\lambda L_{\lambda}}(5100~\AA)}

\newcommand{\llambda}{\lbol$/$\ledd}

\newcommand{\msun}{\ensuremath{M_{\odot}}}
\newcommand{\mbh}{\ensuremath{M_\mathrm{BH}}}

\newcommand{\mgii}{Mg\,{\footnotesize II}}

\newcommand{\nii}{[N\,{\footnotesize II}]}

\newcommand{\oi}{[O\,{\footnotesize I}]}
\newcommand{\oiii}{[O\,{\footnotesize III}]}

\newcommand{\pftest}{\ensuremath{P_{F-{\rm test}}}}

\newcommand{\rosat}{\emph{ROSAT}}

\newcommand{\sii}{[S\,{\footnotesize II}]}

\newcommand{\um}{$\mu$m}

\newcommand{\lax}{{$\mathrel{\hbox{\rlap{\hbox{\lower4pt\hbox{$\sim$}}}\hbox{$<$}}}$}}
\newcommand{\gax}{{$\mathrel{\hbox{\rlap{\hbox{\lower4pt\hbox{$\sim$}}}\hbox{$>$}}}$}}

\accepted{}
\submitjournal{ApJS}

\shorttitle{A catalog of broad-line AGNs} %
\shortauthors{He-Yang~Liu et~al.}

\begin{document}

\title{A COMPREHENSIVE AND UNIFORM SAMPLE OF BROAD-LINE ACTIVE 
GALACTIC NUCLEI FROM THE SDSS DR7}

\correspondingauthor{He-Yang~Liu; Wen-Juan Liu}
\email{liuheyang@nao.cas.cn; wjliu@ynao.ac.cn}

\author{He-Yang~Liu}
\affiliation{Key Laboratory of Space Astronomy and Technology, 
National Astronomical Observatories, Chinese Academy of Sciences, 
20A Datun Road, Chaoyang District, Beijing, 100101, China}
\affiliation{School of Astronomy and Space Science, University 
of Chinese Academy of Sciences, 19A Yuquan Road, Beijing, 100049, China }

\author{Wen-Juan~Liu}
\affiliation{Yunnan Observatories, Chinese Academy of Sciences, 
Kunming, Yunnan 650011, China; Key Laboratory for the Structure and 
Evolution of Celestial Objects, Chinese Academy of Sciences, Kunming, 
Yunnan, 650011, China}

\author{Xiao-Bo~Dong}
\affiliation{Yunnan Observatories, Chinese Academy of Sciences, 
Kunming, Yunnan 650011, China; Key Laboratory for the Structure and 
Evolution of Celestial Objects, Chinese Academy of Sciences, Kunming, 
Yunnan, 650011, China}

\author{Hongyan~Zhou}
\affiliation{Polar Research Institute of China, 451 Jinqiao Road, 
Shanghai, 200136, China}
\affiliation{CAS Key Laboratory for Researches in Galaxies and Cosmology,
University of Sciences and Technology of China, Hefei, Anhui, 230026, China}

\author{Tinggui~Wang}
\affiliation{CAS Key Laboratory for Researches in Galaxies and Cosmology,
University of Sciences and Technology of China, Hefei, Anhui, 230026, China}

\author{Honglin~Lu}
\affiliation{CAS Key Laboratory for Researches in Galaxies and Cosmology,
University of Sciences and Technology of China, Hefei, Anhui, 230026, China}

\author{Weimin~Yuan}
\affiliation{Key Laboratory of Space Astronomy and Technology, 
National Astronomical Observatories, Chinese Academy of Sciences, 
20A Datun Road, Chaoyang District, Beijing, 100101, China}
\affiliation{School of Astronomy and Space Science, University 
of Chinese Academy of Sciences, 19A Yuquan Road, Beijing, 100049, China }
 

\begin{abstract}
A new, complete sample of 14,584 broad-line active galactic nuclei (AGNs) at 
$z<0.35$ is presented, which are uncovered homogeneously from the complete 
database of galaxies and quasars observed spectroscopically in the Sloan Digital 
Sky Survey Seventh Data Release. The stellar continuum is properly removed 
for each spectrum with significant host absorption line features, and careful 
analyses of the emission line spectra, particularly in the \ha\ and \hb\ wavebands, 
are carried out. The broad Balmer emission line, particularly \ha, is used to 
indicate the presence of an AGN. The broad \ha\ lines have luminosities in a 
range of $10^{38.5}$\,--\,$10^{44.3}$~\lum, and line widths (FWHMs) of 
500--34,000~\kms. The virial black hole masses, estimated from the broad-line 
measurements, span a range of $10^{5.1}$\,--\,$10^{10.3}$~\msun, and the 
Eddington ratios vary from $-3.3$ to $1.3$ in logarithmic scale. Other quantities 
such as multiwavelength photometric properties and flags denoting peculiar line 
profiles are also included in this catalog. We describe the construction of this 
catalog and briefly discuss its properties. The catalog is publicly available online. 
This homogeneously selected AGN catalog, along with the accurately measured 
spectral parameters, provides the most updated, largest AGN sample data, which 
will enable further comprehensive investigations of the properties of the AGN 
population in the low-redshift universe.   
\end{abstract}
 

\setcounter{footnote}{0}
\setcounter{section}{0}


\section{INTRODUCTION}

Active galactic nuclei (AGNs\footnote{In this paper, the term ``AGNs'' refers 
to active galactic nuclei at all luminosities, including Seyfert galaxies and quasars. 
Seyfert galaxies are relatively low-luminosity, mostly low-redshift AGNs, whereas 
quasars are generally considered to be the counterparts of Seyfert galaxies at the 
high luminosity regime, e.g., $L_{\rm bol} > 10^{45}$~\lum\ \citep{netzer13}. }) 
refer to a class of energetic phenomena at the central region of galaxies, which 
are believed to be powered by the accretion of galactic material onto supermassive 
black holes (SMBHs). Ever since the recognition of AGNs, including Seyfert 
galaxies \citep{seyfert43} and quasars \citep{schmidt63}, AGNs have become 
one of the most actively studied topics in the astronomy community \citep[e.g,][]
{peterson97,osterbrock06,netzer13}, and enormous effort has been put into 
compiling large samples of AGNs. Large and uniformly selected samples not 
only allow systematic and detailed studies on the phenomenological properties 
of AGNs, but also can yield important insights into the intrinsic physical picture of 
central black holes (BHs) as well as their surrounding galactic environments. In 
addition, it is suggested that most BH mass is accumulated during the luminous 
AGN phases (e.g., \citealp{soltan82,yuqj02,merloni04,shankar09}). Therefore a 
complete census of the AGN population and its properties is particularly 
important to disentangle the BH growth puzzle. In particular, large and well-defined 
samples can help to measure the luminosity function and the BH mass function 
of AGNs, as well as the distribution function of the accretion rates, which are useful 
tools to investigate the AGN demographics and can further provide essential 
constraints on the growth history of SMBHs (e.g., \citealp{ schulze15}). Furthermore, 
the fact that SMBH mass correlates tightly with the mass, luminosity, and velocity 
dispersion of the host bulge \citep[e.g.,][]{magorrian98,ferrarese00,gebhardt00a,
gultekin09}, indicates that there exists a connection between the growth of SMBHs 
and the evolution of their host galaxies. This is also supported by semianalytical 
and numerical simulations on how AGN feedback influences the galaxy formation 
and evolution (e.g., \citealp{dimatteo05,springel05,cattaneo06,croton06,khalatyan08,
booth09}). Large datasets of AGNs, which will drastically extend the physical 
parameter space of the SMBHs (e.g., BH mass, Eddington ratio, spin) and their host 
galaxies (ecoevolution.g., morphology, stellar population, distance, and luminosity), 
can offer unique opportunities to investigate the coevolution between SMBHs and 
their host galaxies.
 
In general, AGNs present observational characteristics different from those of 
stars and normal galaxies, which leads to various methods to search for AGNs. 
Particularly, optical photometric and spectroscopic observations are efficient in 
compiling large AGN samples. As a common practice, color selection in the 
optical band is often used to search for quasar candidates \citep[e.g.,][]
{schmidt83, boyle90} and followed by spectroscopical observations, which are 
essential for identifying their AGN nature and further investigating their 
properties in detail. Systematic study on a well-defined broad-line AGN sample 
based on optical spectroscopy was pioneered by \citet{boroson92}, on the basis 
of  a sample of 87 AGNs with redshifts $z < 0.5$ from the Bright Quasar Survey 
\citep{green86}. Recent progress on large samples of quasars is largely attributed 
to large-scale optical wide-field surveys such as the Sloan Digital Sky Survey 
\citep[SDSS;][]{york00}. The SDSS has opened a new frontier in the study of AGNs 
by providing a huge volume of high-quality spectra of quasars and galaxies. The 
SDSS makes it feasible to select and study in large and homogeneous AGN samples. 
A quasar catalog selected from the SDSS DR7 \citep[DR7Q;][]{schneider10} contains 
105,783 spectroscopically confirmed quasars, which are luminous AGNs ($M_i < 
−22.0$) that exhibit at least one emission line with FWHM $>1000$~\kms, or present 
complex absorption features. \citet{sheny11} conducted spectral measurements 
around the \ha, \hb, \mgii, and \civ\ regions for DR7Q and estimated the BH masses 
using various calibrations. A latest SDSS quasar catalog \citep[DR14Q;][]{paris18} 
derived from the extended Baryon Oscillation Spectroscopic Survey (eBOSS) of the 
SDSS-IV has enlarged the number of known quasars to 526,356.     

Many studies were also performed focusing on selecting AGNs that are fainter 
than typical quasars, namely, the so-called Seyfert galaxies. The optical spectra of 
these underluminous AGNs are generally contaminated by host starlight, thus 
properly modeling and subtracting the stellar components are essential to detect the 
broad-line features of AGNs. To achieve this goal, \citet{haol05} developed a set of 
eigenspectra with main features of absorption line galaxies utilizing the principal 
component analysis \citep[PCA; e.g.,][]{yip04}. They compiled a sample of 1317 
low-redshift AGNs with the FWHM of broad \ha\ $> 1200$~\kms\ from the SDSS 
DR2 \citep{abazajian04}, which was used to investigate the emission line luminosity 
function of AGNs, especially at a low-luminosity regime \citep{haol05b}. Likewise, 
the PCA methods were adopted to deal with the AGN spectra with significant starlight 
in later studies. In an effort to study the relation between AGN luminosity and host 
properties, \citet{vandenberk06} applied the eigenspectrum templates to the 
spectral decomposition of AGNs and host galaxies, yielding a sample of 4666 
low-luminosity type~1 AGNs with FWHM $ > 1000$~\kms\ from the SDSS DR3. 
Using a similar method to deal with the galaxy continuum, \citet{greene07b} built 
a low-redshift broad-line AGN sample from the SDSS DR4 \citep{adelman06}. It 
includes $\sim$8500 objects at $z<0.352$ and was used to investigate the BH 
mass function for broad-line active galaxies in the local universe. Their broad-line 
AGN is defined based on the detection of a broad \ha\ component, and their 
BH masses can extend to the so-called low-mass black hole (LMBH\footnote{As 
described in \citet{liuhy18}, ``Such BHs are also termed intermediate-mass black 
holes (IMBHs) in the literature. We refer to them as LMBHs to avoid confusion 
with ultraluminous X-ray sources (ULXs), which are off-nucleus point-like 
sources, some of which may be powered by IMBHs \citep{kaaret17}''.}) regime 
(\mbh\,$\sim10^6$~\msun). \citet{stern12a} presented a type~1 AGN sample of 
3579 objects using the SDSS DR7 \citep{abazajian09}, which extended the 
DR7Q sample to the low-luminosity end. They used an approach similar to that 
of \citet{haol05} to model the host absorption features and set a threshold on the 
significance of excess flux in the \ha\ region to ensure reliable detection of the 
broad \ha\ line. Many other attempts were also made to select various types of 
AGNs by modeling and subtracting the stellar spectra using different techniques. 
\citet{zhouhy06} selected $\sim$2000 narrow-line Seyfert 1 galaxies (NLS1s) 
utilizing the Ensemble Learning Independent Component Analysis technique 
(EL-ICA; \citealp{luhl06}) to decompose the stellar and nuclear components in 
an accurate way. Using a similar decomposition method, \citet{dongxb12} 
compiled a sample of 8862 sources at $z<0.35$ from the SDSS DR4, which was 
further used to uniformly select LMBH AGNs, particularly those with low 
accretion rates as confirmed by X-ray observations \citep{yuanwm14}. Recently, 
\citet{kyuseok15} identified 1835 new type~1 AGNs with FWHM $ > 800$ \kms\ 
featuring weak broad-line regions (BLRs) at $z < 0.2$ from the SDSS DR7. They 
fit the stellar spectrum by directly matching the observed spectrum with stellar 
templates derived from the \citet{bruzual03} stellar population synthesis models 
and the MILES stellar library \citep{sanchez06}. Their objects have spectra 
predominantly exhibiting host stellar features but also with a weak broad \ha\ 
component.

In previous studies focused on compiling type~1 AGN samples, an operational 
line width \citep[FWHM~$\sim1000$~\kms; e.g.,][]{haol05, schneider10, 
kyuseok15} is often adopted as a demarcation between broad and narrow 
emission lines. This selection criterion works well for luminous AGNs, as the 
emission line width dominated by the broad-line region (BLR) are broad enough 
and thus can effectively exclude those type~2 objects. However, type~1 AGNs in 
the lower-luminosity and lower-BH-mass regimes would fail to meet the FWHM 
$> 1000$~\kms\ criterion. This is particularly true for LMBHs of which the typical 
broad-line FWHM is $\sim$1000~\kms, and the minimum width of their broad 
\ha\ line can even be $\sim500$~\kms\ \citep{ dongxb12, greene07a, liuhy18}. 
These LMBHs with weak broad-line features have been confirmed to be bona 
fide AGNs by their X-ray observations \citep[e.g.,][]{greene07c, desroches09, 
yuanwm14}. Hence, broad-line AGN catalogs selected using the FWHM $> 
1000$~\kms\ criterion would be increasingly incomplete with the decreasing 
AGN luminosity and BH mass. In this study, we do not simply adopt  an 
arbitrary line width as the cutoff between the broad and narrow lines but select 
broad-line AGNs based on the robustness of the broad components in the 
Balmer line regions. Particularly, broad-line AGNs thus selected may be used to 
systematically explore the low-mass end of the local BH mass function (W.-J. Liu 
2020, in preparation).

In this paper, we present a large, uniform and well-defined sample, which 
includes 14,584 broad-line AGNs from the SDSS DR7 at $z < 0.35$. We 
conducted comprehensive spectral analysis and derived complete results of the 
spectral parameters (e.g., line width, line luminosity, etc.), along with a 
compilation of photometric data in a multiwavelength band for each object. As 
described above, the spectra of galaxies and AGNs in the local universe mostly 
show stellar absorption features in their continua that are not negligible, thus it is 
essential to precisely subtract the starlight in order to detect possible broad-line 
features. To model the host continuum, we apply a set of synthesized galaxy 
spectral templates built from the library of simple stellar populations 
\citep[SSP;][]{bruzual03} using the EL-ICA technique developed in our previous 
work \citep{luhl06}.  The robustness and high efficiency of this technique make it ideal 
for modeling and extracting the stellar characteristics for large data sets. In 
addition, in order to obtain accurate emission line measurements, we have 
developed a series of procedures to deal with the decomposition of the coupled 
broad and narrow components in the Balmer line regions, which have been 
applied successfully to the spectral analyses in our previous studies of AGNs 
using the SDSS data \citep[e.g.,][]{dongxb05,dongxb08,dongxb12, zhouhy06,
liuhy18}. The paper is organized as follows. In Section~2, we briefly introduce 
the characteristics of the SDSS, focusing on those aspects most relevant to our 
study. The spectral analysis and sample selection are outlined in Section~3. In 
Section~4, we briefly discuss the sample properties and describe the catalog 
format, followed by a summary in Section~5. Throughout the paper, we assume 
a cosmology with $H_0 = 70$~\kms~Mpc$^{-1}$, $\Omega_m = 0.3$, and $
\Omega_{\Lambda} = 0.7$.

\section{THE SDSS DATA}
Here we briefly summarize on the characteristics of the SDSS\footnote{The 
SDSS DR7 is the final public data release from the SDSS-II occurring in October 
2008, thus here we primarily focus on the characteristics of the SDSS-II. The 
latest updates of the SDSS are available at \url{https://www.sdss.org}.} 
photometry and spectroscopy in this section, and refer readers to 
\citet{abazajian09} for details. The SDSS is a comprehensive imaging and 
spectroscopic survey using a dedicated 2.5\,m telescope \citep{gunn06} located 
at Apache Point Observatory to image over 10,000 ${\rm deg^2}$ of sky and to 
perform follow-up spectroscopic observations. The telescope uses two 
instruments. The first is a wide-field imager \citep{gunn98} with 24 $2048 \times 
2048$ CCDs at the focal plane with 0.396\arcsec\ pixel covering the sky in a 
drift-scan mode in five filters $ugriz$ \citep{fukugita96}. For each filter, the 
effective exposure time is 54.1 s, and 18.75 deg$^2$ are imaged per hour. The 
typical 95\% completeness limits of the images are $u$, $g$, $r$, $i$, $z =$ 
22.0, 22.2, 22.2, 21.3, 20.5, respectively \citep{abazajian04}. The imaging data 
carried out on moonless and cloudless nights of good seeing \citep{hogg01}, are 
calibrated photometrically \citep{tucker06, padmanabhan08} through a series of 
standard pipelines. The photometric calibration is uncertain at the $\sim$1\% 
level in $griz$ and $\sim$2\% in $u$. After astrometric calibration \citep{pier03}, 
the properties of detected objects including the brightnesses, positions and 
shapes are measured in detail \citep{stoughton02}.
  
The second is a 640-fiber-fed pair of multiobjects double spectrographs 
covering the wavelength 3800--9200~\AA\ with a resolution of $\lambda/\Delta 
\lambda$ varying from 1850 to 2200 with a typical signal-to-noise ratio (S/N) of 
10 per pixel for a galaxy near the main sample flux limit. Each optical fiber that 
feeds the spectrograph subtends a diameter of 3\arcsec\ on the sky, 
corresponding to 6.5 kpc at $z = 0.1$. The instrumental dispersion is $
\sim69~\mathrm{km~s^{-1}~pixel^{-1}}$. The targets chosen for spectroscopic 
observations are selected based on their photometric data. Galaxy candidates 
were usually targeted as resolved sources with $r$-band Petrosian magnitudes 
$r < 17.77$, while luminous red ellipticals were specially selected in color--
magnitude space with $r < 19.2$ and $r < 19.5$, respectively, which produce 
two deeper galaxy candidate samples: one is an approximately volume-limited 
sample to $z = 0.38$, and the other is a flux-limited sample extending to $z = 
0.55$ \citep{eisenstein01}. Quasar candidates are selected by their 
nonstellar colors in $ugriz$ broad-band photometry and by their radio emissions 
(unresolved radio counterparts detected by the FIRST survey\footnote{The Faint 
Image of the Radio Sky at Twenty-centimeters (FIRST), is a project designed to 
produce the radio equivalent of the Palomar Observatory Sky Survey over 
10,000 square degrees of the North and South Galactic Caps using the NRAO 
Very Large Array \citep{becker95}.}). This selection targets quasars with $z 
< 3$ to $i$-band point spread function (PSF) magnitude $i = 19.1$ and higher 
redshift quasars ($3 < z < 5.5$) to $i = 20.2$  \citep{richards02}. In addition, 
targeted quasar candidates include optical counterparts to $ROSAT$-detected 
X-ray sources \citep{anderson07}, and other serendipitous objects when free 
fibers were available on a plate. These targets are arranged on tiles of radius 
1.49$^\circ$, and each tile contains 640 objects. The typical spectroscopic 
exposures are 15 minutes, and more time will be taken for faint plates in order to 
reach predefined requirements of S/N, namely, $\rm (S/N)^2$> 15 per 1.5~\AA\ 
pixel for stellar objects of fiber magnitude $g = 20.2$, $r = 20.25$, and $i = 
19.9$. 

The SDSS spectra are well calibrated in wavelength and flux. The flux are 
calibrated by matching the flux of the spectra of standard stars integrated over 
the filter curves to their PSF magnitudes \citep{adelman08}. The flux calibration 
is accurate to 4\% rms, and the wavelength calibration is good to 2~\kms. The 
spectra are classified and redshifts determined by a series of pipelines 
\citep{stoughton02,subbarao02}. The vast majority of these pipeline-determined 
redshifts are reliable with an accuracy of 99\% for galaxies and slightly lower for 
quasars. 
 
\section{SPECTRAL ANALYSIS}
Our aim is to build a complete and well-defined sample of broad-line active 
galaxies in the local universe, based on the SDSS DR7. The SDSS pipeline has 
determined the redshift $z$ and the classification for each spectrum, and the 
accuracy is good enough for this study. We start from the spectra classified as 
``galaxies'' or  ``quasars'' with redshifts below 0.35 to ensure the \ha\ lying within 
the wavelength coverage range of 3800--9200~\AA. This results in a parent 
sample consisting of 854,664 ``galaxies'' and 11,638 ``quasars.'' These spectra 
are first corrected for the Galactic extinction using the extinction map of 
\citet{schlegel98} and the reddening curve of \citet{fitzpatrick99} and are then 
transformed to the rest frame with the redshifts provided by the SDSS pipeline.

There are two key procedures in the spectral analysis:  one is the modeling and 
subtracting of the host stellar continuum, and the other is the decomposition of 
the broad and narrow components in the \ha\ and \hb\ regions. The aperture 
diameter of the SDSS fibers that feed the spectrograph is 3\arcsec\,on the sky, 
which is in general large enough to include  significant starlight from the host 
galaxy in the spectrum. This is particularly true for AGNs fainter than quasars. 
In fact, several interesting subclasses of AGNs, for instance, LMBHs 
\citep[e.g,][]{liuhy18,dongxb12,greene07a}, NLS1s \citep[e.g,][]{zhouhy06}, and 
partially obscured AGNs \citep[e.g,][]{dongxb05}, all show significant starlight in 
their optical spectra.  Hence careful removal of the stellar continuum is necessary, 
which is important for the reliable follow-up measurement of emission lines and 
identification of the broad-line features. For this purpose, we \citep{luhl06} have 
developed a set of starlight templates using the EL-ICA  algorithm, which can be 
used to properly decompose the spectra into stellar and nonstellar nuclear components. 
The details of the continuum fitting method are described in Section~\ref{sec_advfit}.

The broad Balmer emission lines always blend with narrow-line components
and/or the flanking forbidden lines. Thus it is essential to accurately deblend the 
broad Balmer lines from the narrow lines. In general, we adopt the technique
similar to that described in \citet{dongxb12}. The profiles of the narrow lines can 
easily affect the measurement of broad lines, and this can be particularly 
significant for those objects with weak broad-line features. Hence the narrow-line 
profile model is crucial for spectral analysis of those sources at the low-luminosity 
end. We use several different narrow-line profiles to model narrow \ha\ and 
\hb, and the fits with the minimum \chisq\ are adopted as the final result. This is 
physically reasonable, because the narrow Balmer line, for example, \ha, can arise 
from emitting regions with a larger range of critical density and ionization than 
forbidden lines such as \sii, and hence they may have a different profile other 
than \sii\ \citep{ho97b}. A detailed account of the spectral-line fitting technique is 
presented in Section~\ref{sec_linefit}.

For most of the objects, we employ a two-step iteration procedure: first to model 
and subtract the continuum and next to fit the emission lines. Then the emission 
line results are adopted as the initial values of the parameters for the second 
iteration in order to improve the continuum fitting. The iterations are looped until 
the fits of both the continuum and the emission-line spectrum are statistically and 
visually acceptable. For a small fraction of objects (about $\sim$20\% in the final 
sample) showing quasar-like spectra, that is, with negligible starlight contribution 
and prominent and strongly broadened emission lines, a large part of the continuum 
and \feii\ regions would be masked out as emission line regions using the above s
tandard procedure. In addition, the \feii\,$\lambda\lambda$ 4434--4684 multiplets 
are often blended with other broad lines such as \hc, \heii\,$\lambda4686$, and \hb, 
making it complicated to fit the pseudo-continuum in this region. For them, we fit 
simultaneously the continuum together with the emission lines, including the \feii\ 
multiplets, and the forbidden and the Balmer lines. These procedures are detailed 
in the following three absorption line subsections.

\begin{figure*}
\centering
\includegraphics[width=1\textwidth]{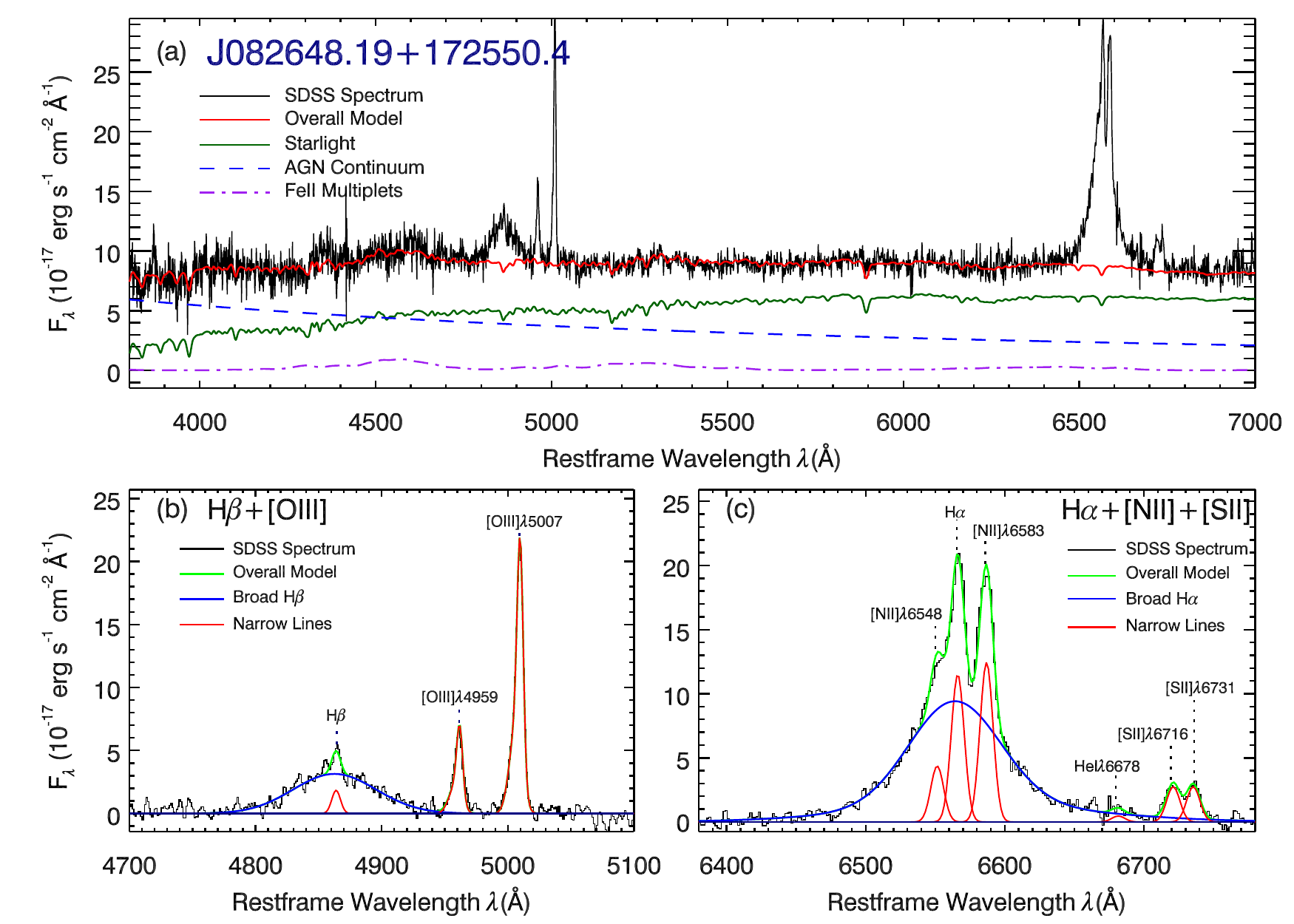}
\caption{\label{fig1}%
Illustration of the continuum and emission line fitting results for an example of 
the broad-line  AGNs with significant stellar contribution in the spectra. This kind 
of spectra is dealt with by the two-step iteration procedure as detailed in the text.
Panel (a): the observed SDSS spectrum (black), the overall model (red), the 
decomposed components of the host galaxy (dark green), the AGN continuum 
(blue dashed), and the \feii\ multiplets (purple dotted-dashed).
Panel (b): emission line profile fit to the \hb~$+$~\oiii\ region.
Panel (c): emission line profile fit to the \ha~$+$~\nii~$+$~\sii\ region.
(The illustration of the continuum and emission-line fits for all the 
SDSS DR7 broad-line AGNs are available at 
\url{http://hea.bao.ac.cn/~hyliu/AGN_catalog.html} and
\url{http://users.ynao.ac.cn/~wjliu/data.html}.)
}
\end{figure*}

\subsection{Starlight and Continuum Subtraction}
\label{sec_advfit}
The vast majority of the parent SDSS sample galaxies ($\sim$99\%) have stellar 
absorption features in their spectra because they are predominately normal galaxies. 
Hence, proper starlight templates are essential for the continuum fitting. One rough 
approach is to use the spectra of absorption line galaxies as templates to 
directly fit the emission-line-free regions of galaxy spectra \citep[e.g.,][]{filippenko88}. 
However, it is difficult and time-consuming to find a proper template because the 
spectra of absorption line and emission line galaxies are quite different. Many 
attempts have been made to improve the stellar templates in the past decades. 
PCA is one of these efforts. \citet{haol05} applied the PCA technique to a sample 
of pure absorption line galaxies and obtained a stellar template library composed 
of eight eigenspectra. PCA is a statistical method that converts the data into a set 
of orthogonal and linearly uncorrelated principal components (PCs): the first PC 
(PC1) contains the maximum variance of the data; PC2 is the second best explanation 
of the variability of the data and must be orthogonal to PC1; and so forth for each 
succeeding PC. When applied to stellar template construction, PCA can greatly 
reduce the number of the galaxy templates but include most of the stellar 
features in the original galaxy spectra library. This can significantly cut down the 
computational time cost of the fit and is ideal to deal with large data sets. The 
ICA algorithm adopted in this study is a blind separation technique that can transform 
the multidimensional statistical data to a set of statistically independent components 
(ICs). ICA can be considered as an extension of PCA and inherits its strength of 
convenience and low time cost in spectral fits. They both attempt to find a set of 
vectors for the original data so that each point in the data can be described as a 
linear combination of these vectors. However, ICA is more powerful. PCA requires 
their PCs to be linearly uncorrelated and orthogonal, while ICA is based on the 
assumption that the ICs are statistically independent with non-Gaussian distributions 
but do not need to be orthogonal. Independence is much more concrete than 
uncorrelatedness, and orthogonality is invalid for a general mixing process. Thus 
ICA can obtain underlying information from the data when PCA fails. The EL 
algorithm is a low-cost method that approximates an unknown intractable posterior 
distribution $p$ by a tractable separable distribution $q$ qualified by a relative 
entropy $d(q||p)$\footnote{Entropy is a measure of uncertainty in the information 
theory, and relative entropy can be used to estimate the `distance' between two 
distributions. Particularly, the Kullback-Leibler distance is adopted to measure 
the relative entropy in \citet{luhl06}.}. EL can be applied to ICA and helps to find 
the simplest function for the ICs as an interpretation of the data\footnote{According 
to the well-known Ockham's razor principle, simpler solutions are more likely to be 
correct than complex ones.}, which thus can greatly mitigate overfit. We \citep{luhl06} 
applied the EL-ICA technique to a library of 1326 stellar spectra\footnote {Note that 
only a subsample of 74 `significantly different' spectra was used in the EL-ICA 
analysis because this algorithm converges very slowly. The ICs derived from the 
subsample have been used to fit the 1326 SSP spectra and the results show that 
they can well represent the parent sample. The criteria for `significantly different' are 
detailed in \citet{luhl06}.} spanning a wide range of ages and metallicities \citep{bruzual03}, 
and a set of six nonnegative six ICs, which can explain 97.6\% of the 1326 SSP 
spectra, are chosen as the final galaxy templates. The spectra of these ICs are 
visually distinctive from each other and imply a tight correlation between stellar 
population age and the ICs. In general, the galaxy templates constructed using 
the EL-ICA  method can be used to fit a spectrum of any stellar system with a 
low computational cost and thus are suitable for dealing with large data sets. 

As described above, the nuclear emissions from AGNs, especially 
low-luminosity Seyfert galaxies, are often contaminated by the stellar absorption 
lines of the host galaxy. Thus before AGN selection, we first model and subtract 
the stellar continua from the SDSS spectra. A pseudo-continuum model, which
takes both the contribution from the host galaxy and the nucleus into account, is 
adopted to fit the continuum. The so-called pseudo-continuum is a nonnegative 
linear combination of several properly generated templates, including the stellar 
component, nuclear continuum, and the optical \feii\ multiplets. The Balmer 
continua and high-order Balmer emission lines are added if they can improve the 
fit by decreasing the reduced \chisq\ by at least 20\%. The starlight component is 
modeled by a combination of six synthesized galaxy templates broadened by 
convolving with a Gaussian of width to match the stellar velocity dispersion of 
the host galaxy. These synthesized galaxy templates, as detailed above, 
are built up from the spectral template library of SSPs \citep{bruzual03} using 
the EL-ICA algorithm \citep{luhl06} and have included the vast majority of the 
stellar features in SSPs and can greatly reduce the chance of overfit. Particularly, 
we slightly shift the starlight model in adaptive steps to correct for the effect 
brought by the possible uncertainty of the redshifts provided by the SDSS pipeline, 
and the fit with the minimum reduced \chisq\ is adopted as the final result. This 
can help to measure the host galaxy velocity dispersion and to detect weak 
emission lines more accurately.  A reddened power law of  $f(\lambda) \sim 
\lambda^{-1.7}$ \citep{francis96}, with $E(B-V)$ varying from 0 to 2.5 assuming 
an extinction curve of the Small Magellanic Cloud \citep{peiyc92}, is adopted 
to model the AGN continuum emission. The optical \feii\ multiplets are modeled 
with two separate sets of templates in analytical forms, one for broad lines and 
the other for narrow lines. These \feii\ templates are constructed based on the 
measurements of I\,Zw\,1 by \citet{veron04}; see \citet{dongxb08,dongxb11} for 
details. The merit of the analytical-form \feii\ templates is that they enable us to 
fit the \feii\ multiplets with any line width. This is particularly useful in the case 
of LMBHs, where the spectra tend to show significantly narrower \feii\ than those 
of I\,Zw\,1 \citep[see][]{greene07a}.  

During the pseudo-continuum fitting, two kinds of spectral regions are masked 
out: (1) bad pixels as flagged by the SDSS pipeline and (2) wavelength ranges 
emission line that may be seriously affected by prominent emission lines. A 
composite SDSS quasar spectrum introduced by \citet{vandenberk01} is initially 
adopted to determine the second mask region. After modeling and subtracting the 
continuum, the leftover pure emission line spectrum is fitted using the method 
described in Section~\ref{sec_linefit}. Then the measured emission line 
spectrum is used to replace the composite quasar spectrum in the next iteration 
of the continuum fitting. This two-step procedure is reiterated until that both the 
continuum and the emission line fitting results are statistically acceptable (either 
the reduced \chisq\ $< 1.1$ or the fit cannot be significantly improved with a 
chance probability less than 0.05 according to the $F$-test, \pftest$< 0.05$\footnote{
\label{fn-ftest} The null hypothesis that model~2 does not provide a significantly 
better fit than model~1 is rejected if the $P$ is lower than the critical value of 
$F$-distribution (e.g., $P = 0.05$ ), which corresponds to the probability 
of incorrectly rejecting a true null hypothesis. In general, $P = 0.05$ corresponds 
to a typical false-rejection probability of 50\%, and the false-rejection probability 
decreases with lower $P$.}). An illustration of the starlight-nucleus decomposition 
is shown in Figure~\ref{fig1}.

\begin{figure*}
\centering
\includegraphics[width=1.0\textwidth]{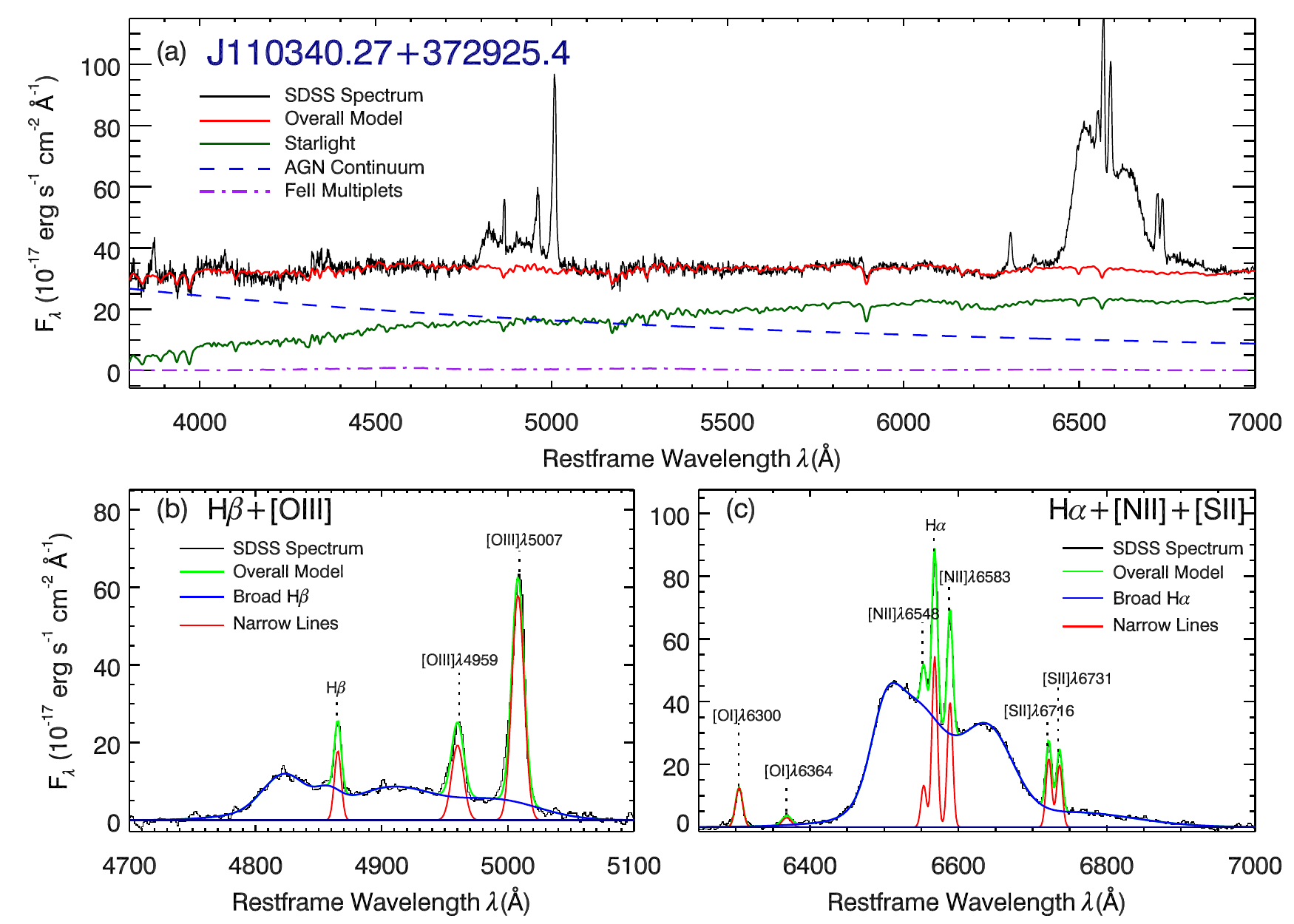}
\caption{\label{fig2}%
An example of AGNs with complicated broad line profiles. The symbols and lines 
are the same as those in Figure~\ref{fig1}. 
}
\end{figure*}

\subsection{Emission Line Fitting}
\label{sec_linefit}

The emission line fitting procedure adopted in this work is based on and improves 
upon that used in \citet{dongxb05} and \citet{dongxb12}, which is summarized 
below. We primarily focus on two emission line regions, namely, \ha\ + \nii\ (or 
\ha\ + \nii\ + \sii\  if the line width of broad \ha\ is very large) and \hb\ + \oiii. The 
\ha\ + \nii\ region is generally complicated because the possible broad \ha\ always 
blend with the narrow \ha\ and the flanking \nii\ doublets. Thus it is necessary to 
impose some constraints on the narrow lines in order to reduce the number of 
free parameters. A common approach is to model and fix the profiles of narrow 
\ha\ and \nii\  using that of \sii, because empirically the line profile of \sii\,$
\lambda\lambda6716, 6731$ is generally well matched to those of \nii\,$
\lambda\lambda6548,6583$ and narrow \ha. This has been confirmed by 
\citet{zhouhy06} using a sample of $\sim$3000 type~2 AGNs selected from the 
SDSS, which shows that both the line widths of both \nii\ and \sii\ are statistically well  
consistent with that of narrow \ha. In some cases, \sii\ is undetectable or too 
weak to yield a reliable measurement, and the \oiii\ profile is used as a 
substitute. The \oiii\ emission lines seem to be significant in almost all the 
spectra with possible broad-line features. However, they often present 
an asymmetrical profile and  have broad, blue shoulders, which are suggested to 
originate from outflows. Thus the centroid profile of \oiii\,$\lambda$5007 is 
adopted if \oiii\ is fitted using multiple Gaussians. In general, our fitting strategies 
for emission lines are as follows:

1.  Generally, each narrow line is fitted with a single Gaussian. More Gaussians 
are added if the \sii\ or \oiii\ doublets  show complex profiles and are detected 
with S/N $> 10$ (up to 3 for \oiii\ and 2 for \sii, respectively). The multi-gaussian 
scheme is adopted if it improves the fit by  \pftest $< 0.05$. The \sii\  doublets 
are assumed to have the same profile and fixed in separation by their laboratory 
wavelengths. The same is applied to the \oiii\,$\lambda\lambda$4959,5007 doublet 
lines. In addition, the flux ratio of the \oiii\ doublets $\lambda5007/\lambda4959$ 
is fixed at the theoretical value of 2.98. 

2. The narrow \ha\ and \nii\,$\lambda\lambda6548,6583$ are assumed to have 
the same profiles and redshifts as the \sii\ doublets, or the core of \oiii\,$
\lambda\lambda4959,5007$ if \sii\ is absent or weak. The profiles and redshifts 
of \ha\ and the \nii\ doublets are assumed to be the same as \sii\ or \oiii. Similar 
to the \oiii\  doublets, the flux ratio of the \nii\ doublets $\lambda6583 /
\lambda6548 $ is fixed to be 2.96.

3. If the broad \hb\ is not detected and the narrow \hb\ can be well fitted with one 
Gaussian, the narrow \ha\ will be given the same profile as the narrow \hb.  

4. If the broad and narrow \hb\ components cannot be separated by the fit, then 
the centroid wavelength and profile of the narrow \hb\ will be tied to those of \ha, 
\oiii, or \sii, in this order. 

5. The broad \ha\ line is usually fitted using one or two Gaussians, and the broad 
\hb\ is assumed to have the same redshift and profile as the broad \ha\ or set to 
be free if it can significantly improve the fit of the \hb\ + \oiii\ region by \pftest $< 
0.05$. In fact, the line widths of broad \hb\ are systematically slightly larger than 
those of broad \ha\ \citep[e.g.,][]{greene05b}, hence the profiles of broad \hb\ are 
set to be free in a moderate fraction of the final broad-line sample. In some 
spectra, the broad Balmer lines show asymmetric or irregular profiles such as 
double or even more velocity peaks, and multiple Gaussians (up to 5)  are 
adopted in these cases (see Figure~\ref{fig2}). 

\begin{figure*}
\centering
\includegraphics[width=1.0\textwidth]{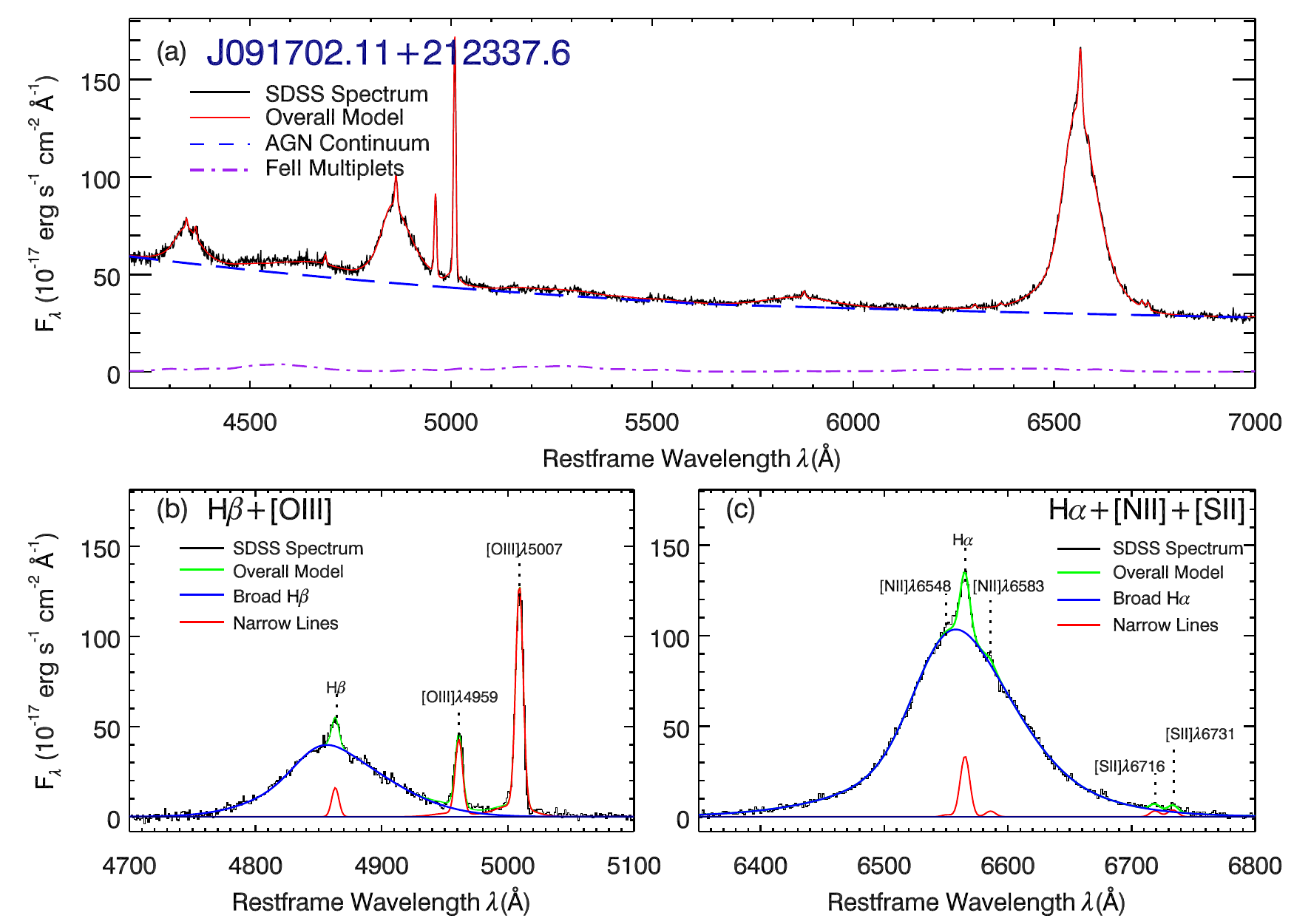}
\caption{\label{fig3}%
A representative example of AGN spectra without starlight contamination, for 
which simultaneous fitting of the continuum and emission lines is conducted. 
Panel (a): the observed SDSS spectrum (black), the overall model (red), 
the AGN continuum (blue dashed), and the \feii\ multiplets (purple 
dotted-dashed).
Panel (b): emission line profile fit to the \hb~$+$~\oiii\ region.
Panel (c): emission line profile fit to the \ha~$+$~\nii~$+$~\sii\ region.
}
\end{figure*}

The emission line fitting procedure is composed of  three steps. First, we fit the 
\ha\ + \nii\ region using pure narrow-line profiles without a broad component. The 
narrow \ha\ component and the \nii\,$\lambda\lambda6548,6583$ doublets are 
modeled and fixed using the \sii, \oiii, or narrow \hb, in this order. In the second 
step, an additional component is added for each spectrum to account for 
possible broad \ha\ if the \chisq\ decreases significantly with \pftest $< 0.05$. 
Thus a subsample of broad-line candidates is selected from the parent sample 
based on the new added broad \ha\ components that are requested to 
conform to the criteria that (1) FWHM of the broad component is relatively larger 
than those of narrow lines, particularly \oiii\,$\lambda5007$, and (2) flux of the 
broad \ha\ is statistically significant, namely, greater than the flux error by a factor 
of 3, and meanwhile greater than $10^{-16}$~\flux.

Finally, we apply the refined fitting procedure for each candidate broad-line 
source to improve the accuracy of the line measurement. The profiles of the 
narrow emission lines in AGNs vary with critical density and ionization 
parameters. Narrow Balmer lines can arise from emitting regions that may 
have different physical conditions from those of forbidden-line emitters. Thus the 
profile of the narrow \ha\ or \hb\ may be different from that of \sii\ or \oiii.  
Considering this fact, we employ several different schemes to model the narrow 
\ha. The narrow \ha\ is fitted using (1) a model built from the best-fit \sii\ with one 
or double Gaussians; (2) a single-Gaussian model from the best-fit core 
component of \oiii; (3) a multiple-Gaussian model from the best-fit global profile 
of \oiii; and (4) a single-Gaussian model built from narrow \hb, if broad \hb\ is not 
detected and narrow \hb\ can be fitted well with one Gaussian, respectively. In 
each scheme, the fitting is constrained following the strategies described above. 
The fitting results of these different schemes are compared with each other, and 
the one with the minimum \chisq\ is adopted as the final result. 

\subsection{Simultaneous Fit of Continuum and Emission lines}
As mentioned above, for objects showing quasar-like spectra, we fit 
simultaneously the nuclear continuum and the \feii\ multiplets, together with other 
emission lines (see Figure~\ref{fig3}). The method is similar to that 
described in \citet{dongxb08}, but with some modifications to improve the 
accuracy of the measurement of the broad components of \ha\ and \hb, which 
is described as follows. Each spectrum is fitted in the rest-frame wavelength 
range of 4200$-$7200~\AA\ using a combination of nuclear continuum, 
\feii\ multiplets, and prominent emission lines in the \ha\ and \hb\ regions. 
The nuclear continuum is modeled by a broken power law with a break 
wavelength of 5650~\AA, for example, $a_1\lambda^{-\alpha_{\lambda,1}}$ 
for the \hb\ region and $a_2\lambda^{-\alpha_{\lambda,2}}$ for the \ha\ region. 
The break wavelength of 5650~\AA\ is adopted because it can ideally avoid the 
wavelength regions of the prominent emission lines based on the AGN 
composite spectrum given by \citet{vandenberk01}. The optical \feii\ emission is 
modeled as $C(\lambda) = c_b C_b(\lambda)+c_n C_n(\lambda)$, where 
$C_b(\lambda)$ and $C_n(\lambda)$ represent the broad and narrow \feii\ line
templates in analytical form constructed using the measurements of I Zw~1 
provided by \citet{veron04}, as in their Table A1 and A2, respectively. The 
emission lines other than iron lines are modeled following the strategy 
described in Section~\ref{sec_linefit}. 

\begin{figure*}
\centering
\includegraphics[width=1.1\textwidth, trim=40 10 20 0,clip]{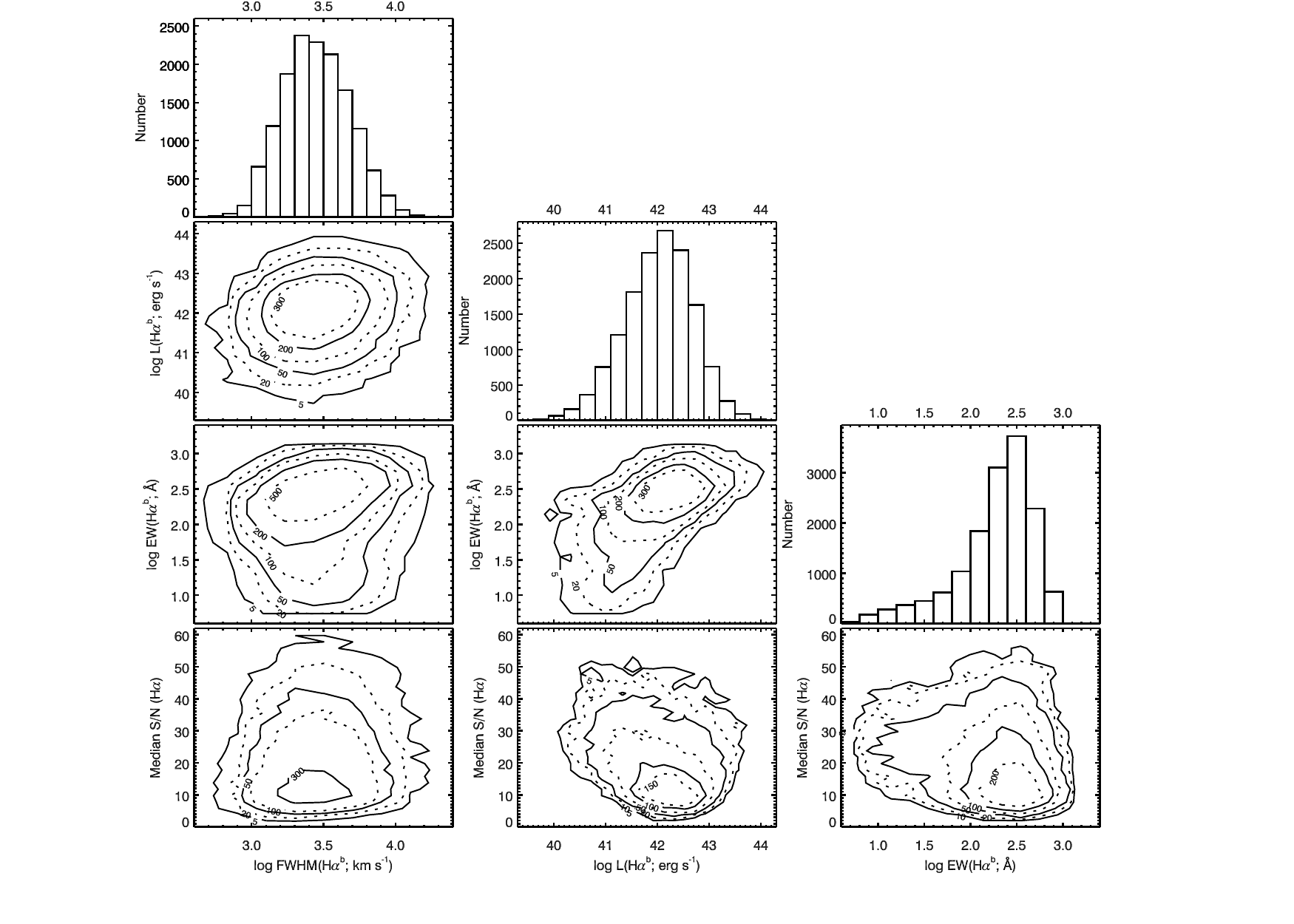}
\caption{\label{fig4}%
Distributions of the line luminosity, FWHM, EW, and S/N of broad \ha\ line for 
the present DR7 broad-line AGN sample objects. Contours are local point 
density contours, and the values of the contour levels are marked for each 
isoline (the same in the following contour plots). 
}
\end{figure*}

To ensure the reliability of the line measurements, we recalculate the reduced 
\chisq\ of the fits around the \ha\ and the \hb\ regions, respectively, and spectra 
with reduced \chisq\ $>1.1$ in either region are picked out for further refined 
fitting. The refined fitting procedures are similar to those described in 
Section~\ref{sec_linefit}. Briefly, we fit each pseudo-continuum-subtracted 
spectrum using various narrow-line models, and the one with the minimum 
reduced \chisq\ is adopted. The broad components are fitted using up to five 
Gaussians; the results are accepted when the reduced \chisq\ $<1.1$ or it 
cannot be improved significantly by adding in one more Gaussian with a chance 
probability $<0.05$ according to the $F$-test.

\subsection{Broad-line Selection Criteria}
The conventionally broad line criterion of FWHM\,$\sim1000$~\kms\ \citep[e.g.,]
[]{haol05,schneider10,kyuseok15} would inevitably reject a large fraction of 
type~1 AGNs at the low-mass and low-luminosity end, and hence results in 
significant incompleteness of the sample. In this work, we adopt the selection 
criteria following \citet{dongxb12} and \citet{liuhy18}, based directly on the 
presence of a broad component (c.f. the narrow lines) from the results of 
emission line fitting, particularly, broad \ha, which is generally the strongest 
broad line in the optical spectra of AGNs. The criteria are as follows: 

\indent 1. $P_{F\rm-test} < 0.05$, \\
\indent 2. Flux(\bha) $> 10^{-16}$ \flux, \\
\indent 3. S/N(\bha) $\geq 5$,  \\
\indent 4. $h_{\rm B} \geq 2$\,rms, and \\
\indent 5. FWHM(\bha) $>$ FWHM(NL),

\noindent
where $P_{F{\rm -test}}$ is the chance probability given by the $F$-test and is 
used to test whether adding a broad \ha\ component in the pure narrow-line 
model can significantly improve the fit or not; S/N(\bha) $=$ Flux(\bha)$/
\sigma_{\rm total}$, \bha\ denotes the broad component of \ha\ line, 
$\sigma_{\rm total}$ is the total  uncertainty of broad \ha\ 
arising from statistical noise, continuum subtraction, and subtraction of nearby 
narrow lines, as will be elaborated below; $h_{\rm B}$ is the height of the 
best-fit broad \ha\ component, and rms is the quadratic mean deviation of the 
continuum-subtracted spectra in the emission-line--free region near \ha; FWHM 
is full width at half maximum, and NL refers to narrow line. %

\begin{figure*}
\centering
\includegraphics[width=1.1\textwidth, trim=40 10 20 0,clip]{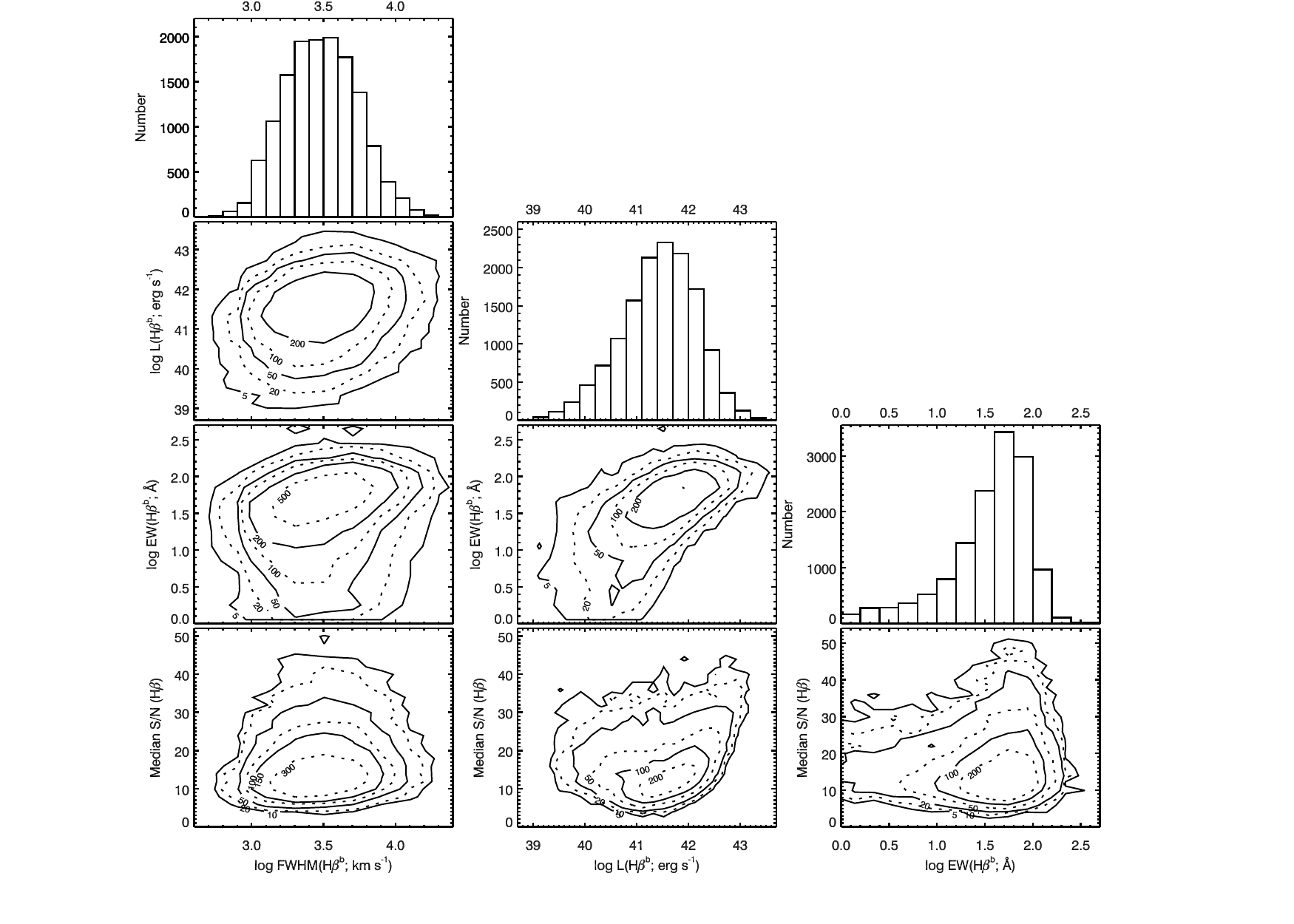}
\caption{\label{fig5}%
Distributions of the line luminosity, FWHM, EW, and S/N of the broad \hb\ line for 
the present DR7 broad-line AGN sample objects. }
\end{figure*}

Criterion~1 gives the statistical significance of detecting a broad \ha\ given by 
the $F$-test. We have found through simulations that these criteria based on the 
$F$-test work well, although theoretically, these goodness-of-fit tests hold only 
for linear models (see \citealp{haol05}). The flux limit in criterion~2  is set to 
eliminate possible spurious detections caused by underlying inappropriate 
continuum subtraction. The broad \ha\ component detected at $10^{-16}$~\flux\, 
appears to be a fake broad feature because a slight fluctuation in the continuum 
may result in a marginal broad \ha\ detection with such a flux. In fact, even the 
minimum broad \ha\ flux in our final broad-line AGN sample is much larger than 
this threshold.  Criterion~3 is to ensure the reliability of the existence of the 
broad component in terms of S/N. S/N(\bha) $=$ Flux(\bha)$/\sigma_{\rm total}$, 
and $\sigma_{\rm total}$ is the quadrature sum of statistical noise ($\sigma_{\rm stat}
$), the uncertainties arising from the subtraction of continuum ($\sigma_{\rm 
cont\_sub}$) and narrow lines ($\sigma_{\rm NL\_sub}$), namely, $
\sigma^2_{\rm total} = \sigma^2_{\rm stat} + \sigma^2_{\rm NL\_sub} + 
\sigma^2_{\rm cont\_sub}$~. The $\sigma_{\rm stat}$ is given by the fitting code 
MPFIT. The significance of the $\sigma_{\rm NL\_sub}$ depends on the width 
and intensity of the broad \ha\ line. In cases where broad \ha\ are broad and 
strong, the decomposition between broad and narrow lines is apparent and has 
little influence on the measured flux of broad \ha; this term is small and 
negligible. However, for a broad \ha\ line that is both narrow and weak, this term 
may be significant, because the line deblending is highly dependent on the 
narrow-line model. Hence we estimate the $\sigma_{\rm NL\_sub}$ based on fitting 
results using different narrow models. In the stage of refined fitting for emission 
lines, we have employed several different narrow-line models to deal with each 
spectrum. The rest $n$ sets of fitting results that are worse than the best one 
with a chance probability of the $F$-test greater than 0.1 are selected to 
estimate the $\sigma_{\rm NL\_sub}$, 
\begin{equation}
\footnotesize
\sigma_{\rm NL\_sub} = \sqrt{\frac{ \sum_{i=1}^{n} ({\rm Flux(\bha})_i - 
{\rm Flux(\bha)}_{\rm best})^2}{n}}~.
\end{equation}
\noindent
As for $\sigma_{\rm cont\_sub}$, in practice, this term is difficult  to estimate. 
The \ha\ absorption lines in those spectra with old stellar populations may 
moderately affect the flux measurements. However, this situation is rare in our 
final sample. On the whole, we suppose the error term caused by the continuum 
subtraction is negligible because the \ha\ absorption features are weak in most 
cases and the flux uncertainty of broad \ha\ introduced by continuum 
decomposition is much below 1~$\sigma$ (also see the discussion in 
footnote~12 of \citealp{dongxb12}). Criterion~4 is a supplementary constraint, 
which can help to minimize spurious detections mimicked by narrow-line wings 
or fluctuation in the continuum. In the vast majority of cases, the selected 
broad-line objects based on the S/N threshold in criterion~3 have $h_{\rm B}$ 
much larger than 2~rms. Criterion~5 is set to ensure the ``broad'' feature of the 
broad component in terms of line width. It requires that the FWHM of the broad 
\ha\ be larger than those of narrow lines. Our analysis yielded a total of 14,584 
sources (with duplicates removed\footnote{We also removed two objects 
identified as supernovae: SDSS J132053.66+215510.2 \citep{Izotov09} and 
SDSS J140102.50+452433.7 (H.Y. Zhou 2019, private communication). }) that 
pass these criteria. 

\begin{figure}
\centering
\includegraphics[width=0.48\textwidth]{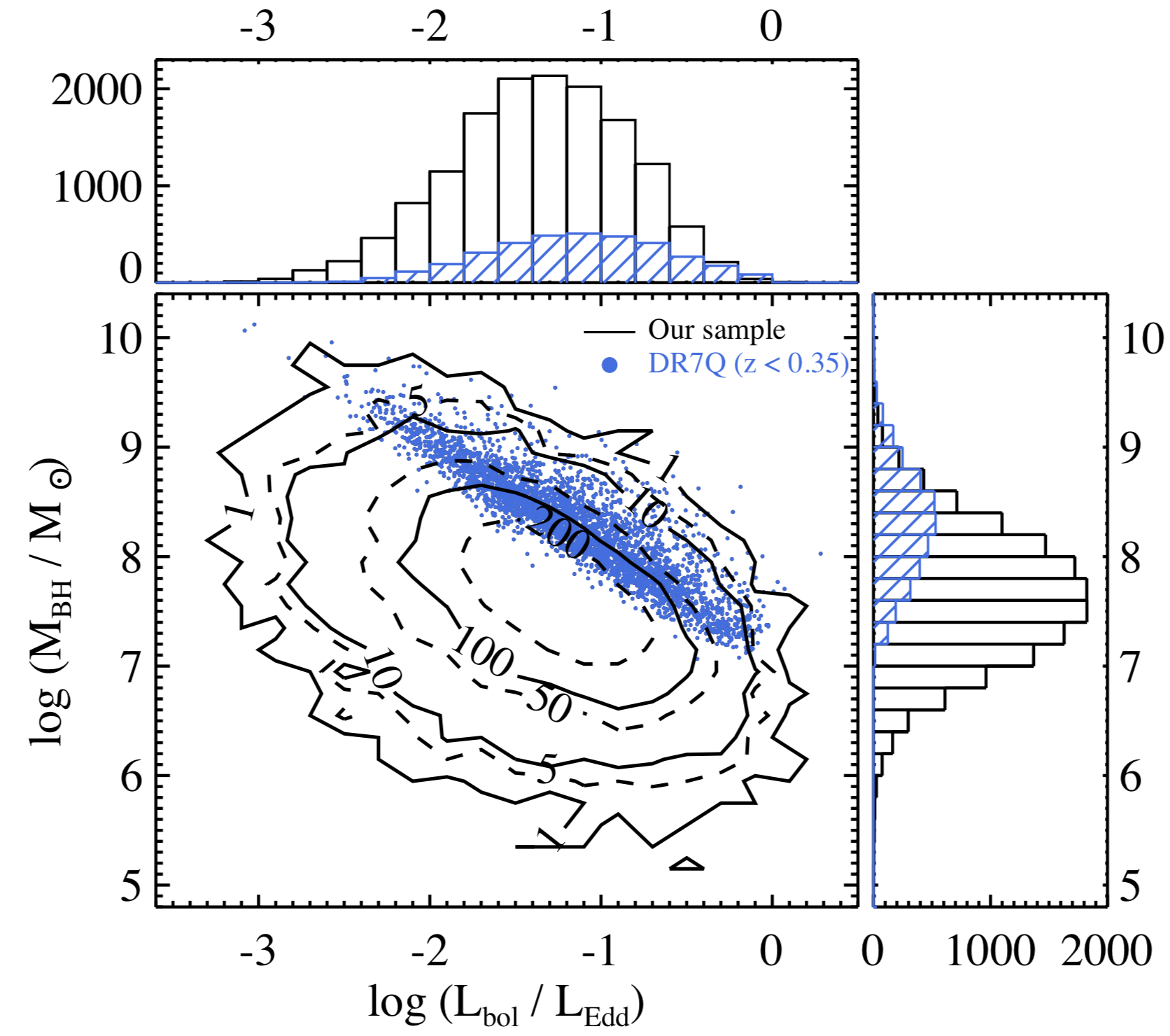}
\caption{\label{fig6}%
Distributions of the BH mass and Eddington ratio for our broad-line AGN 
sample (black contour) and the DR7Q sample at $z<0.35$ \citep[blue circles,][]
{schneider10,sheny11}.  The deficit of data points at the upper right corner 
indicates that there exist few fast accreting massive AGNs in the local universe, 
while the absence of objects at the lower left region is due to the flux limit of the 
SDSS. The right and top panels show the BH mass and Eddington ratio 
histograms for our broad-line AGNs (black histograms), respectively. The blue 
hatched histograms represent the distribution of the DR7Q objects at $z<0.35$. 
The DR7Q terminates at \mbh~$\approx10^7$~\msun, while the BH masses of 
the broad-line AGNs in our sample span a wide range of 
$10^{5.1}-10^{10.3}$~\msun.
}
\end{figure}

\section{THE SAMPLE}
\subsection{Sample Properties}
We have compiled a sample of 14,584 type~1 AGNs based on detection of a 
broad \ha\ line from a total of about one million SDSS DR7 spectroscopic 
objects. The median $z$ of the final broad-line AGN sample is 0.2. The line 
luminosities, FWHMs, and EWs are measured for the entire sample. The 
FWHMs are measured from the multi-Gaussian models as described in 
Section~\ref{sec_linefit}, and their uncertainties are derived from 
simulations\footnote{A series of fake emission line spectra are generated using 
the statistical uncertainties given by the MPFIT, and the uncertainty of FWHM is 
estimated from these generated fake spectra.}. Figure~\ref{fig4} and 
Figure~\ref{fig5} show the distributions of these quantities for broad \ha\ and \hb. 
The FWHMs of \bha\ span a range of $\sim$500$-$34000\footnote{Only one 
object SDSS J094215.12+090015.8 has broad \ha\ with FWHM of 34000~\kms, 
and the FWHM(\bha) of the others are located in the range of $\sim$500$-
$22000~\kms.}~\kms, with a median of 2760~\kms. The broad \ha\ line 
luminosities \lbha\ reside in the range of  $\sim10^{38.5}-10^{44.3}$~\lum\ with a 
median lying at $10^{42.1}$ \lum. EWs of broad \ha\ cover a range of 
$5-900$~\AA\ with a median of 232~\AA. For the broad \hb\ line, the FWHMs, 
luminosities, and EWs are located in the range of $\sim500-$34000~\kms, 
$10^{38.9}-10^{43.7}$~\lum, and $\sim1-780$~\AA, respectively; and their 
medians are 3020~\kms, $10^{41.5}$~\lum, and 43~\AA, respectively. The 
distributions of the median S/N per pixel around the \ha\ and \hb\ regions show 
that most of the emission line measurements except for line luminosity have no 
significant dependence on the S/N, which indicates that the selection of broad-line 
AGNs is unbiased by the S/N of the spectra. As for the dependence of line 
luminosity on S/N, we consider that it is more likely to be caused by the selection 
effect due to the SDSS flux limit.   

\begin{figure*}
\centering
\includegraphics[width=1.0\textwidth]{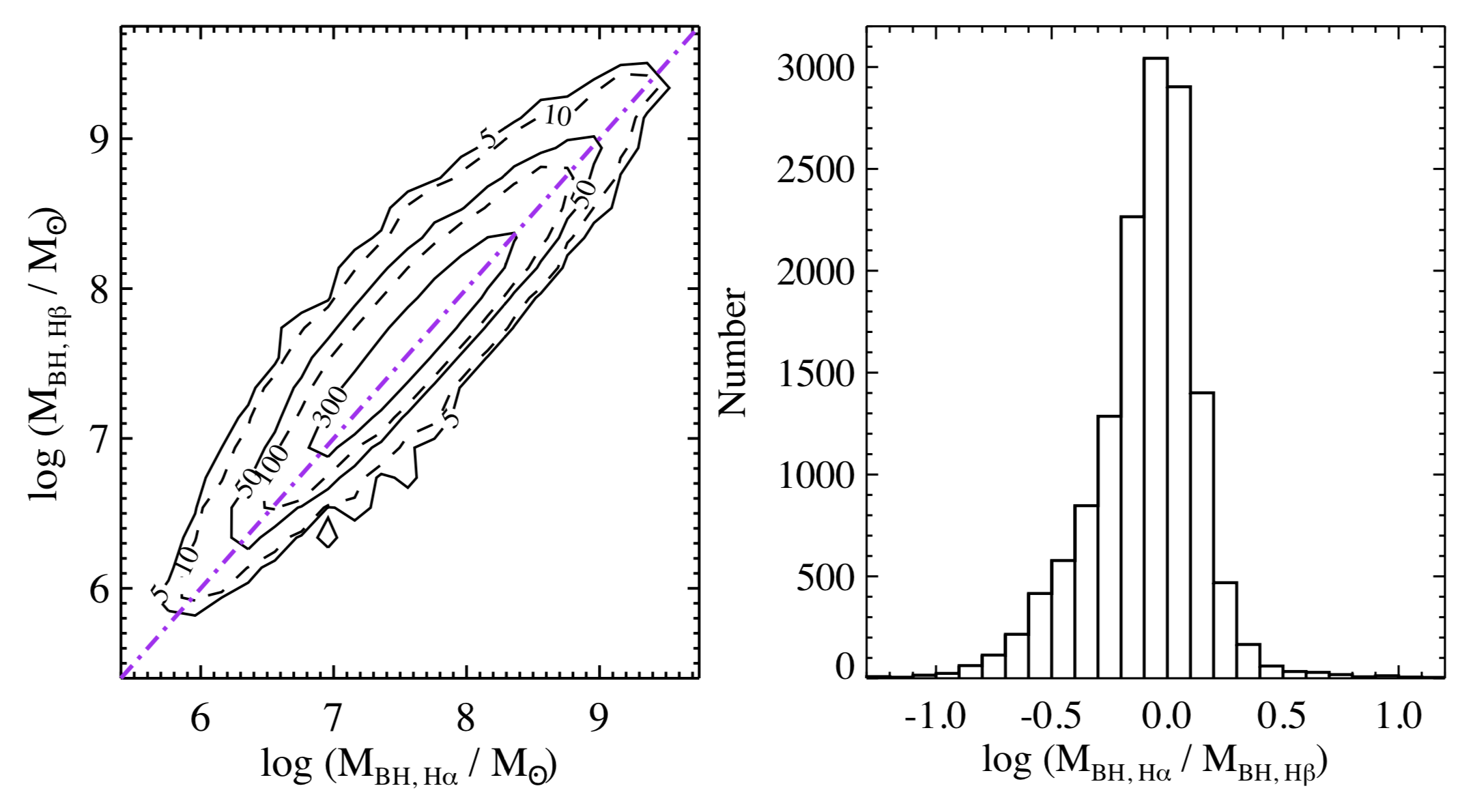}
\caption{\label{fig7}%
Comparison of virial masses derived from two different line estimators based on 
the broad \ha\ and \hb\ lines, respectively, for our broad-line AGN sample objects
of which both line estimates are available. The contour in the left panel 
presents their distribution on the $M_{\rm BH,H\alpha}$ versus  $M_{\rm 
BH,H\beta}$ space, and the marked values on the isolines are corresponding 
number densities. The dotted-dashed line denotes 1:1. The right panel shows 
the distribution of mass ratios between these two lines. The BH masses derived 
from broad \hb\ are slightly larger than those obtained from broad \ha.   
}
\end{figure*}

The BH masses can be estimated using the spectral measurements based on 
the virial method. This method assumes that the BLR system is virialized; the 
BLR radius is estimated from the continuum luminosity using the radius-luminosity 
(R-L) relation derived from reverberation mapping (RM) studies of AGNs 
\citep[e.g.,][]{kaspi05,bentz06,bentz09a,bentz13,wangjg09}, and the virial 
velocity is indicated by the broad-line width (FWHM or line dispersion). The 
scaling factor depends on the structure of the BLR system and can be 
calibrated by other BH mass estimators such as $M-\sigma_{*}$ relation 
(e.g., \citealp{ferrarese00,gebhardt00a}). Here we adopt the BH estimator 
outlined in \citet{ho15}, which was calibrated using a larger and more current 
sample of reverberation mapping AGNs than previous studies and taking 
into account the recent determination of the virial coefficient for pseudo and 
classical bulges. This estimator is subject to an intrinsic scatter of 0.35~dex. 
The \mbh\ based on broad \hb\ are derived as follows:
\begin{equation}
\footnotesize
\rm log{\it M}_{BH} = log\left[\left(\frac{FWHM(\hb)}{1000~\kms}\right)^2\left(
\frac{L_{5100}}{10^{44}\lum}\right)^{0.533}\right] + 6.91,
\end{equation}
\noindent
where $L_{5100}$ is the continuum luminosity at rest frame 5100~\AA\ 
($L_{5100} \equiv$ \lfive), which is derived from the AGN continua decomposed 
from the spectra or from the broad \hb/\ha\ luminosity using the scaling relation 
given in \citet{greene05b} if the spectra are dominated by starlight. \mbh\  are 
also estimated based on the broad \ha\ using the above formalism and the 
empirical correlation between FWHM of broad \ha\ and \hb\ given by 
\citet{greene05b}. The final nominal \mbh\ for each object is adopted as the 
mean of \mbh(\hb) and \mbh(\ha) whenever both are possible. 
The BH masses span a range of $10^{5.1}-10^{10.3}$~\msun, with a median 
of $10^{7.6}$~\msun. The Eddington ratio (\llambda) is calculated using the 
measured $L_{5100}$ and the nominal BH masses. It is defined as the ratio 
of the bolometric luminosity ($L_{\rm bol}$) to the Eddington luminosity 
($\ledd  = 1.26 \times 10^{38} \mbh/\msun$). $L_{\rm bol}$ is derived from the 
optical continuum luminosity at 5100~\AA\ using a bolometric correction factor 
of 9.8 \citep {mclure04}, $L_{\rm bol}=9.8L_{5100}$. The Eddington ratios thus 
estimated range from $-$3.3 to 1.3 in logarithmic scale, with a median of $-$1.3. 
Figure~\ref{fig6} shows the distributions of BH masses and Eddington ratios. 

\begin{figure*}
\includegraphics[width=1.0\textwidth]{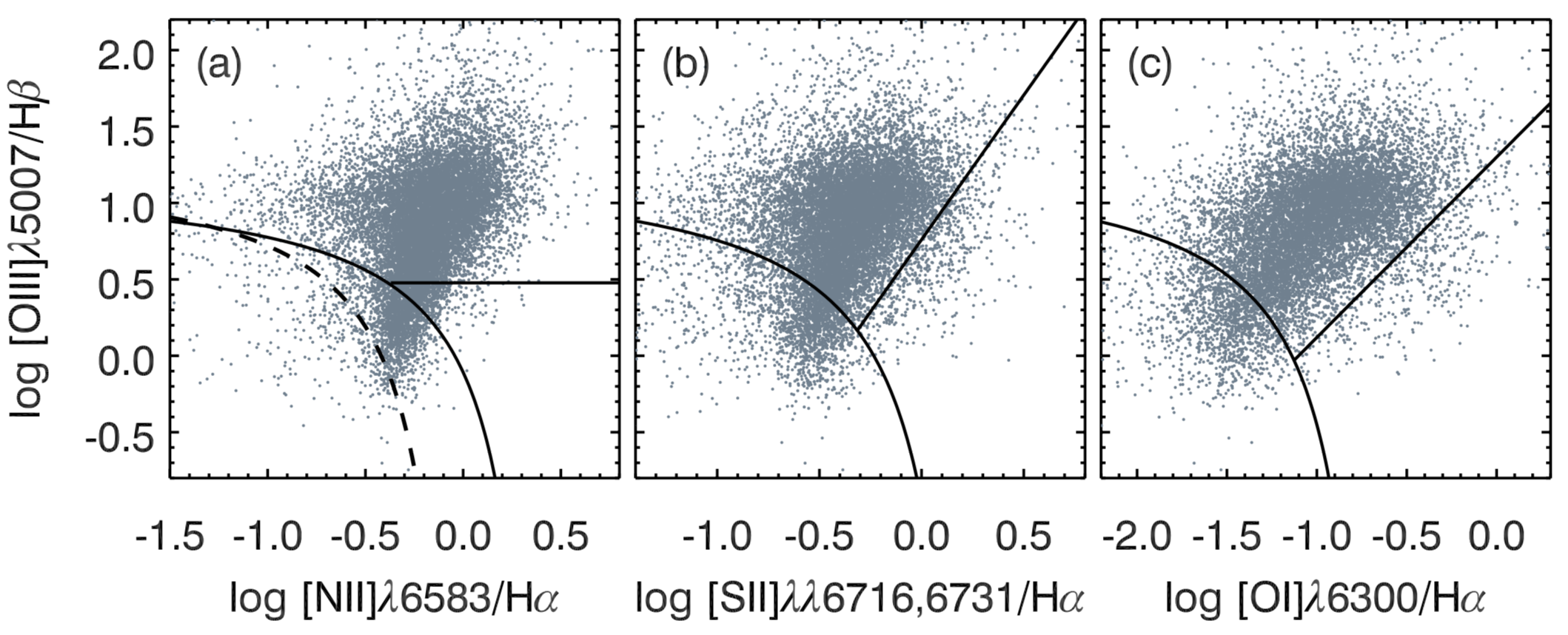}
\caption{\label{fig8}%
Narrow-line diagnostic diagrams of \oiii\,$\lambda 5007/$\hb\ versus \nii\,$
\lambda 6583/$\ha\ ({\it a}), versus \sii\,$\lambda\lambda 6716,6731/$\ha\ ({\it 
b}), and versus \oi\,$\lambda 6300/$\ha\ ({\it c}) for the broad-line AGN sample in 
this study. The extreme starburst classification line (solid curve from 
\citealp{kewley01} and the Seyfert--LINER line (solid straight line in panel $b$ 
and $c$) obtained by \citet{kewley06} are adopted to separate the \hii\ regions, 
AGNs and LINERs. In panel ($a$), the dashed line corresponds to the pure star 
formation line given by \citet{kauffmann03b}, and the horizontal line represents 
\oiii\,$\lambda 5007/$\hb\ $=3$, which is conventionally used to separate Seyfert 
galaxies and LINERs.}
\end{figure*}

Figure~\ref{fig7} presents the comparison between the BH 
masses estimated from \ha\ and \hb. The BH masses derived from broad \hb\ 
are slightly larger than those obtained from broad \hb. The direct reason is that 
the \ha-based BH mass formalism is not calibrated independently, but by 
substituting into the \hb-based BH mass estimator using the empirical FWHM 
relation between broad \ha\ and broad \hb\ (Eq.~3 of \citealp{greene05b}). 
The intrinsic mechanism is complicated; in fact, there are a lot of systematics 
involved in deriving these \ha- or \hb-based BH masses. For instance, the local 
R-L relation is biased toward the most variable AGNs \citep{bentz13}, thus 
derived BH mass estimators using single-epoch spectra suffer large uncertainties; 
the relative profile of \ha\ and \hb\ depends on both the kinematic and ionization 
structure of the BLR \citep{osterbrock06}, which can vary quite significantly among 
different objects; dust extinction can also alter the profile of Balmer emission 
lines, and the Balmer decrements for the objects in our sample span a wide range. 
Many ($\sim$68\%) are larger than the typical \ha/\hb\ of 3.5 for objects in \citet{greene05b}.  

\begin{figure}
\centering
\includegraphics[width=0.48\textwidth]{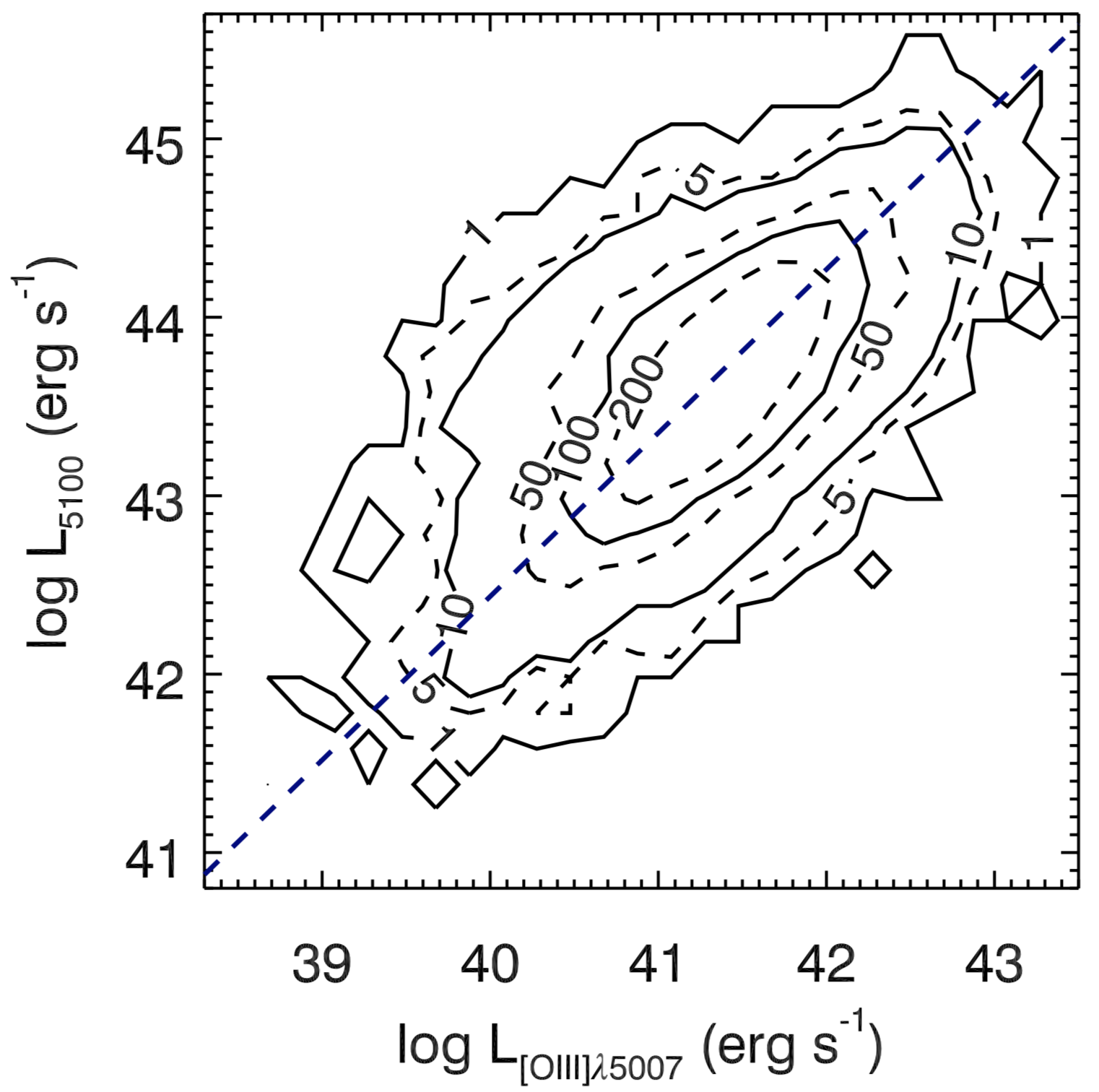}
\caption{\label{fig9}%
Correlation between the \oiii\ line luminosity ($L_{\rm \oiii\,\lambda5007}$) and 
the continuum luminosity at 5100~\AA\ ($L_{5100}$). The navy dashed line 
denotes a bisector linear regression fit described in the text, and the best-fit 
relation is $ {\rm log}(L_{5100}) = 0.92{\rm log}(L_{\rm \oiii\,\lambda5007})
+5.76$, with a scatter of $\sim$0.43~dex.  
}
\end{figure}

AGNs can also be identified from their loci in the diagnostic diagrams 
constructed by the relative strengths of various prominent narrow emission lines, 
namely, so-called BPT diagrams, which are commonly used to select narrow-line 
AGNs. The BPT diagrams involving the narrow-line ratios of \ha, \hb, \oiii, \nii, 
\sii, and \oi\ are shown in Figure~\ref{fig8}. The vast majority of our type~1 AGNs 
are located in the region of either Seyfert galaxies or composite objects. Only a 
small fraction (3\%) fall into the pure star-forming region, which can be explained 
by the large aperture of 3\arcsec\,of the SDSS that inevitably includes light from 
the host galaxies. This result confirms the AGN nature of our sample. Moreover, 
the AGN nature was partly verified by the X-ray properties of some of these 
broad-line objects, which is particularly important, or those showing weak 
broad-line features such as LMBHs \citep{yuanwm14,liuhy18}.

Figure~\ref{fig9} shows the correlation between the \oiii\,$\lambda5007$ 
luminosity and the continuum luminosity at 5100~\AA. This scaling relation is 
generally used to estimate the bolometric luminosity using the \oiii\,$
\lambda5007$ luminosity as a proxy for type~2 AGNs (e.g.,\citealp 
{kauffmann03b, heckman04, reyes08}. However, as noted in previous studies 
(e.g., \citealp{heckman04,reyes08}), this correlation suffers from a large scatter. 
The \oiii\,$\lambda5007$ line is expected to be affected by interstellar dust 
extinction and the complicated condition makes it difficult to correct. This would 
introduce uncertainty in this correlation as well. Here we give a rough linear 
relation between $L_{\rm \oiii\,\lambda5007}$ and $L_{5100}$, namely, $ {\rm 
log}(L_{5100}) = 0.92{\rm log}(L_{\rm \oiii\,\lambda5007}) +5.76$, with a scatter 
of $\sim$0.43~dex.

\begin{figure*}
\includegraphics[width=1.0\textwidth]{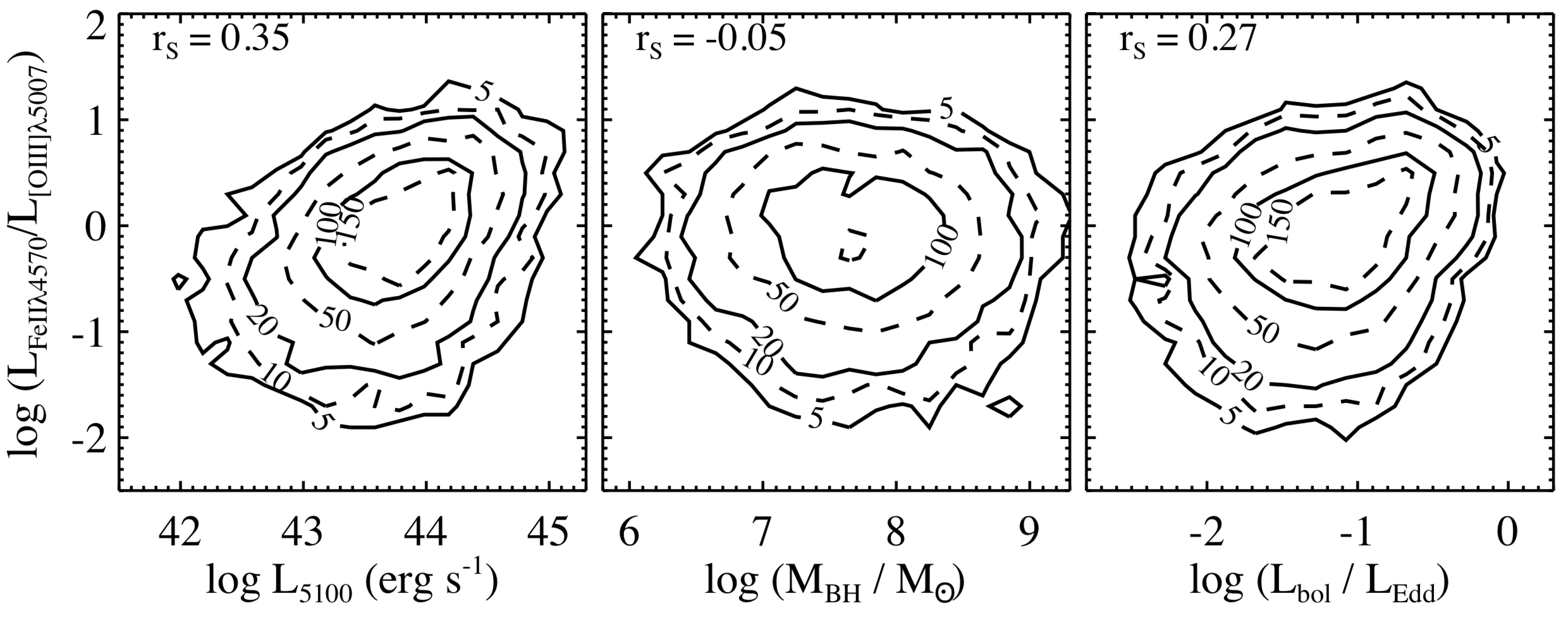}
\caption{\label{fig10}%
Dependence of the line ratio of \feii\,$\lambda4570$ to \oiii\,$\lambda5007$ 
on the continuum luminosity $L_{5100}$, BH mass (\mbh ), and Eddington ratio 
(\llambda), respectively. Also denoted are the Spearman correlation coefficient 
$r_{\rm S}$. The relative strength of \feii\,$\lambda4570$ to \oiii\,$
\lambda5007$ shows moderate dependence on $L_{5100}$ and \llambda\ but 
no correlation with \mbh.
}
\end{figure*}

Figure~\ref{fig10} shows the dependence of the intensity ratio \feii\,$
\lambda4570$ to \oiii\,$\lambda5007$ ($L_{\feii\,\lambda4570}/L_{\rm \oiii\,
\lambda5007}$) on $L_{5100}$, \mbh, \llambda, which is in the framework of the 
eigenvector~1 AGN parameter space introduced by \citet[EV1 or PC1;][]
{boroson92}.  The flux of the \feii\,$\lambda4570$ is integrated in the rest-frame 
wavelength range 4434--4684~\AA. EV1 was suggested to be physically driven 
by the relative accretion rate or the Eddington ratio \citep[e.g.,][]{boroson92, 
boroson02}. This was confirmed by \citet{dongxb11} utilizing a large, 
homogenous AGN sample selected from the SDSS, who found that the intensity 
ratio of \feii\,$\lambda4570$ to \oiii\,$\lambda5007$ indeed correlates most 
strongly with \llambda, rather than with \mbh\ or $L_{5100}$.  However, in this 
work, we find that $L_{\feii\,\lambda4570}/L_{\rm \oiii\,\lambda5007}$ is 
correlated with both $L_{5100}$ and \llambda\ at similar levels (with Spearman's 
rank coefficients of $r_{\rm S} = 0.35$, 0.27, and chance probabilities of $p = 
9 \times 10^{-37}$, $5\times 10^{-31}$, respectively). No correlation is found with 
\mbh\ ($r_{\rm S} = -0.05$ and $p = 0.003$). The physical mechanisms driving 
these correlations in the parameter space are currently not clear and deserve 
further study in the future.

\subsection{Comparisons with Other Broad-line AGN Samples}
\subsubsection{Comparison with the SDSS DR7 quasar sample} 
\begin{figure}
\centering
\includegraphics[width=0.48\textwidth]{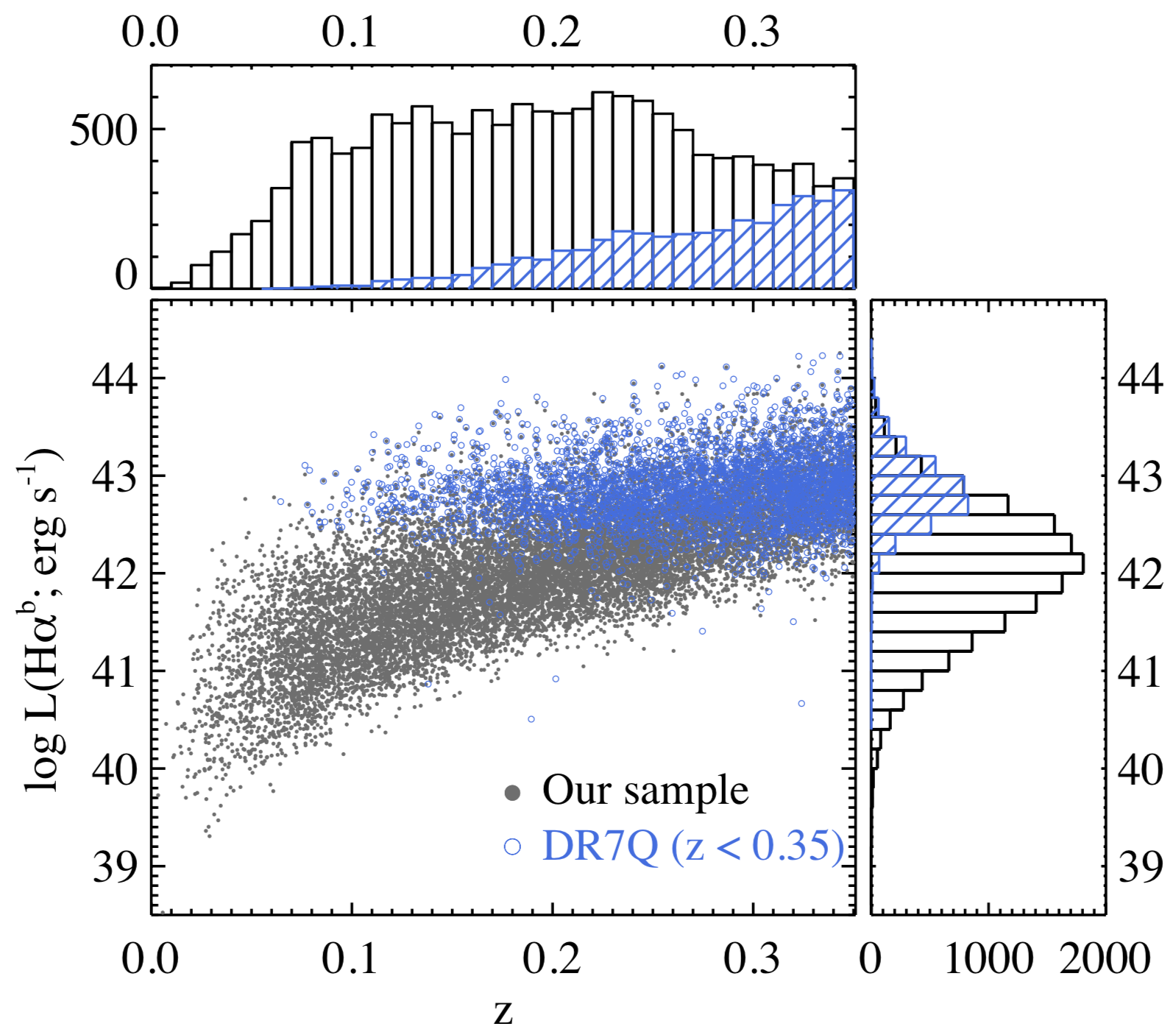}
\caption{\label{fig11}%
Distributions of type~1 AGNs in this study (gray filled circles) and DR7Q 
objects at $z<0.35$ (blue open circles) on the \lbha\ versus $z$ plane. The loci 
of these two samples overlap at $z > 0.3$, and our sample becomes distinct due 
to the lack of faint AGNs in DR7Q at $z < 0.3$. The right and top panels show 
the luminosity and redshift histograms, respectively. Note that the vast majority 
of DR7Q objects at $z<0.35$ reside at \lbha\,$> 10^{42}$~\lum, while our sample 
extends further down to \lbha\,$\approx 10^{39}$~\lum.
}
\end{figure}

The quasar catalog based upon the SDSS DR7 (DR7Q, \citealp{schneider10,
sheny11}), which contains 105,783 objects with luminosities higher than
$M_i = -22.0$ and redshifts at $0.065<z<5.46$, was selected using the classical 
broad-line criterion of FWHM $> 1000$~\kms. It is interesting to compare the 
current sample with DR7Q because our study adopts a different strategy, which is 
based directly on the detection of broad Balmer lines for the selection of type~1 
AGNs. Figure~\ref{fig11} presents the distributions on the \lbha\ versus $z$ plane 
for DR7Q and our sample. The loci of these two samples coincide at $z>0.3$; 
however, our sample becomes distinct at $z<0.3$ due to the deficit of 
low-luminosity AGNs in DR7Q. DR7Q includes 3532 objects at $z < 0.35$, 
3419 out of which are also included in our sample. The right panel in Figure~\ref{fig11} 
compares the distribution of \lbha\ spanned by our sample with DR7Q at $z < 
0.35$. The median \lbha\ of DR7Q  at $z < 0.35$ is 42.9 in logarithmic scale, 
systematically higher than the 42.1 of our sample; in particular, our sample includes 
6733 objects at \lbha $<10^{42}$~\lum, in sharp contrast with 25 in DR7Q. The 
distributions of FHWM(\bha) for DR7Q at $z < 0.35$ and for our sample are 
generally consistent within 0.06~dex (their medians are 3150~\kms\ and 
2760~\kms, respectively). Furthermore, the DR7Q objects at  $z<0.35$ have 
BH masses\footnote{Note that the virial BH masses in DR7Q are not estimated 
based on broad \ha; for details see the Section 3.7 in \citet{sheny11}.} spanning 
from $10^{7.1}$~\msun\ to $10^{10.5}$~\msun\ with a median of 
$10^{8.3}$~\msun, while our sample extends down by almost 2 orders in 
magnitude in \mbh\ and includes 2154 objects with \mbh\ below $10^7$~\msun\ 
(see Figure~\ref{fig6}). In terms of the Eddington ratios, the difference between 
these two samples is not so significant as in BH mass; the median \lratio\ of our 
sample is about 0.2~dex lower than that of DR7Q ($-$1.35 vs. $-$1.13). In 
summary, our sample is a good complement to DR7Q, in the sense that it enlarges
the number of low-$z$ type~1 AGNs by a factor of 4 (thus more complete) and 
extends the SDSS AGN sample to the lower-luminosity and lower-BH-mass end. 
   
The present sample, which puts more efforts into finding faint AGNs, provides a 
large database of local AGNs spanning wide ranges of various quantities, 
for example, luminosity, line width, BH mass, Eddington ratio, and so on, especially 
in the low-BH-mass regime. This can help to investigate the properties of AGNs 
of particular interest such as low-luminosity AGNs with low Eddington ratios 
(LINERs),  active BHs at the lowest-mass regime (LMBHs), and NLS1s that are 
suggested to be accreting at their Eddington limits. In particular, the LMBHs in this 
sample can help study the connection between SMBHs and their host galaxies in the 
low-mass regime, which is poorly explored so far. LMBHs are suggested to be 
predominately located in pseudobulges, which seem not to  follow the correlations 
between SMBHs and the quantities of classical bulges (e.g, mass, luminosity, 
velocity dispersion). In addition, being well defined and more complete in the 
low-BH-mass regime than previous ones, our sample can be used to construct 
the BH mass function and luminosity function for BHs in the local universe. 
These studies, especially for AGNs with \mbh~$<10^6$~\msun, can provide
important constraints for models of the formation and evolution of BHs 
\citep{greene07b,volonteri10}.  

\subsubsection{Comparisons with other low-luminosity type~1 AGN samples}
\citet{greene07b} constructed a sample of $\sim$8500 broad-line active galaxies 
(hereafter GH07) from the SDSS DR~4 with redshift $z<0.352$, of which 92\% 
(7839) are included in our sample as well. The broad-line detection rates of 
GH07 and our samples are comparable; this is not surprising because the 
broad-line selection strategies of these two samples are similar. The medians of 
redshift, broad \ha\ luminosity, and FWHM of GH07 sample are 0.2, 42.06, and 
2830, respectively, which are similar to those of our sample (the differences are 
within 0.1~dex). Note that our sample have more objects located in the 
low-luminosity regime (the fraction of sources with \lbha~$<10^{41}$~erg 
s$^{-1}$ in our sample is 8.1\% vs. 4.5\% of the GH07 sample), which indicates 
that our approach has the advantage of detecting fainter AGNs with weaker 
broad \ha, as demonstrated in \citet{dongxb12} and \citet{liuhy18}  as well (LMBH 
AGN samples).

\citet{stern12a} obtained a sample of 3579 type~1 AGNs with $z<0.31$ 
(hereafter SL12), selected also from the SDSS DR~7 based on the detection 
of broad \ha\ emission. Their sample objects have a broad \ha\ luminosity of 
$10^{40}-10^{44}$~erg~s$^{-1}$ with a median of $10^{42}$~erg~s$^{-1}$, 
and a broad \ha\ width (FWHM) of 1000--25000~\kms\ with a median of 
3200~\kms. The vast majority of the objects in the SL12 sample (3468) are 
included in our sample as well, while we note that the SL12 sample is much 
smaller than ours. The absence of a large fraction of the broad-line AGNs in our 
sample in SL12 may be due to their stricter criteria on selecting the broad \ha 
line: the FWHM of broad \ha\ larger than 1000~\kms\ and the broad \ha\ flux 
larger than four times the scatters of the continuum-subtracted spectrum near \ha. 
These two further criteria excluded AGNs with narrower or weaker broad lines. 
In addition, they applied several filters on the DR7 database, which resulted in 
a substantial decrease in the size of the parent sample. 

\subsection{Description of the Catalog}
We have tabulated the measured quantities from the spectral fitting for the 
14,584 broad-line AGNs, along with their photometric data in multiwavelength 
bands. This catalog\footnote{A fits version of this catalog can be accessed at 
\url{http://hea.bao.ac.cn/~hyliu/AGN_catalog.html} and \url{http://users.ynao.ac.cn/~wjliu/data.html}.} 
is composed of three parts: the first contains the basic 
information of each object; the second part is the emission line measurements 
derived from the spectral analysis in this work; the third part is a compilation of 
the multi-wavelength photometric measurements obtained from other surveys. 

\input{table1.tex}

\subsubsection{Object Information}
Part~I of the catalog (see Table~\ref{tab1}) is composed of the basic information 
of these sources including the identification number assigned in this paper, the 
SDSS name, coordinates, redshift, spectroscopic plate, modified Julian date 
(MJD) and fiberid, together with the number of spectroscopic observations. The 
objects in this catalog are sorted by their R.A. and Dec, and their redshifts are 
taken from the SDSS pipeline. Plate is an integer indicating the SDSS plug plate 
that was used to collect the spectrum;  MJD denotes the modified Julian date 
of the night when the observation was carried out; fiberid denotes the fiber id 
(1--640 for SDSS DR7). Any given spectrum can be identified by the 
combination of these three numbers. In the SDSS, spectra for many objects are 
taken simultaneously, and thus some objects were observed more than once 
with different plate-MJD-fiberid combinations. For those objects with multiple 
observations, we adopt the spectrum of the highest median S/N and record the 
number of spectroscopic observations in the flag $N_{\rm spec}$.

\input{table2.tex}

\subsubsection{Emission line Measurements}
The measured quantities obtained from the spectral analysis in this work are 
included in the second part, and the format is described in Table~\ref{tab2}. Here 
we briefly describe the specifics of the cataloged quantities. 
These line fluxes are observed values and are not corrected for intrinsic 
extinction or dust reddening. The line FWHM and luminosity that can be inferred 
from other lines are not included, for example, \nii\,$\lambda$6548, whose FWHM and 
luminosity can be obtained from those of \nii\,$\lambda$6583.  The flux of \feii\,$
\lambda4570$ is calculated by integrating the flux density of the corresponding 
\feii\ multiplets from 4434~\AA\ to 4684~\AA\ in the rest frame. We do not give 
the uncertainty arising from the statistical measurement for \mbh, because the 
uncertainty of the virial BH mass estimator is dominated by systematic effects 
(0.3\,--\,0.5~dex; e.g., \citealp{greene06,wangjg09,grier13}). Quantities for those 
undetectable lines are set to be $-$999. The format of the line measurement 
catalog is as follows.

\begin{enumerate}
\item[1.] Identification number assigned in this paper.

\item[2-9.]  Centroid wavelength, line luminosity, FWHM, rest-frame equivalent 
width, and their uncertainties for the broad \ha\ component.

\item[10-17.]  Centroid wavelength, line luminosity, FWHM, rest-frame equivalent 
width, and their uncertainties for the narrow \ha\ component.

\item[18-21.] Line luminosity, rest-frame equivalent width, and their uncertainties 
for \nii$\lambda$6583.

\item[22-25.] Line luminosity, rest-frame equivalent width, and their uncertainties 
for \sii$\lambda$6716. 

\item[26-29.] Line luminosity, rest-frame equivalent width, and their uncertainties 
for \sii$\lambda$6731.

\item[30-33.] Line luminosity, rest-frame equivalent width, and their uncertainties 
for \oi$\lambda$6300.

\item[34-41.]  Centroid wavelength, line luminosity, FWHM, rest-frame 
equivalent width, and their uncertainties for the broad \hb\ component.

\item[42-46.]  Centroid wavelength, together with line luminosity, FWHM, 
rest-frame equivalent width, and their uncertainties for the narrow \hb\ component.

\item[47-54.]  Centroid wavelength, line luminosity, FWHM, rest-frame 
equivalent width, and their uncertainties for \oiii$\lambda$5007.

\item[55-58.] Line luminosity, rest-frame equivalent width, and their uncertainties 
for \feii$\lambda$4570.

\item[59-60.] $L_{5100}$ and its uncertainty measured by the power-law 
continuum or broad \ha\ from the spectral fits.

\item[61.] This flag denotes if the emission lines show multiple peak profiles: 0 
$=$ no emission line with multiple peaks; 1 $=$  \oiii$\lambda$5007 with multiple 
peak profile; 2 = broad Balmer line with multiple peak profile.

\item[62.] Virial black hole mass derived based on \ha.

\item[63.] Virial black hole mass derived based on \hb.

\item[64.] The adopted fiducial virial black hole mass. The fiducial virial black 
hole mass is obtained from the mean of Item~62 and Item~63.

\item[65.] Eddington ratio derived from the fiducial virial BH mass.

\end{enumerate}

\subsubsection{Multiwavelength Photometric Measurements}
In Part~III of the catalog, we supplement photometric measurements in the 
optical, ultraviolet (UV), near-infrared (NIR), mid-infrared (MIR), X-ray and 
radio, and the format is detailed in Table~\ref{tab3}. In the optical band, we 
provide both the PSF (point spread function) and Petrosian magnitudes in SDSS 
$ugriz$ for these objects, given that our sample is a mixture of quasars and 
Seyfert galaxies. The PSF magnitude can well describe the luminosity of a pointed 
source with aperture correction that can effectively mimic the dependence of 
measured magnitude on the aperture radius and seeing variations. The 
Petrosian magnitude is defined as the flux within a consistent aperture radius 
\citep{petrosian76,blanton01} and can give a good measurement for the 
photometry of nearby galaxies. The photometric measurements in other bands 
are drawn from GALEX, 2MASS, $WISE$, \rosat, and FIRST. The specific 
details of these surveys are described below.

We collect photometry data in the UV band from the archive of the $Galaxy Evolution 
Explorer$ mission ($GALEX$; \citealp{martin05}), which has imaged over two-thirds 
of the sky in two bands: far-ultraviolet (FUV; $\lambda_{\rm eff}=1528$~\AA) and 
near-ultraviolet (NUV; $\lambda_{\rm eff}=2271$~\AA). Its FUV detector stopped 
working in 2009 May, and the subsequent observations have only NUV data. The 
latest data release is GR~7\footnote{The GALEX GR~7 can be accessed at the 
MAST website \url{http://galex.stsci.edu/GR6/}.}. Here we use the revised 
catalog constructed by \citet{bianchi17}, which includes all the sources detected 
by GALEX based on the latest GR~6 and GR~7 and duplicate measurements of 
the same objects have been properly removed. We search for the UV 
counterpart of our AGNs with a matching radius of 5\arcsec\ \citep[e.g.,][]
{morrissey07, stern12a}\footnote{In general, we adopt a commonly used 
matching radius in previous studies.}. A subsample of 11,551 objects are found 
to have GALEX detections.

The NIR photometric information is obtained from the Two Micron All-Sky Survey 
(2MASS; \citealp{skrutskie06}). The 2MASS is an all-sky near-infrared survey in 
the $J$ (1.25~\um), $H$ (1.65~\um), and $K_{\rm s}$ (2.16~\um) photometric 
bands with a positional accuracy of 0.5\arcsec\, and photometric accuracy of $
\sim$5\% for bright sources. The 2MASS includes two catalogs: the Point 
Source Catalog (PSC) and the Extended Source Catalog (XSC). Candidate 
point sources are identified and subtracted from images by the pipeline, and 
eventually included in the PSC. The XSC contains sources that are extended 
with respect to the instantaneous PSF, such as galaxies and Galactic nebulae. 
Here we search for a 2MASS counterpart for each object in our sample from 
both PSC and XSC, using a matching radius of 2\arcsec\,and 6\arcsec, 
respectively \citep[e.g.,][]{skrutskie06,stern12a}; both results are included. The 
PSC magnitude was measured in standard aperture with a radius of 4\arcsec, 
and a curve-of-growth correction was applied to compensate possible light loss 
\citep{skrutskie06}. Note that unlike the PSC, the XSC does not provide a
``default'' magnitude; we present the photometric data based on the 
extrapolation of the surface brightness profile that reflect the total fluxes of the 
sources\footnote{ We caution that the `extrapolation' photometry is vulnerable to 
surface brightness irregularities and thus estimated color is not good enough in 
some cases. The elliptical isophotal photometry is a better choice to estimate 
accurate colors for galaxies.}.  

The $Wide-field Infrared Survey Explorer$ ($WISE$; \citealp{wright10}) provides a 
sensitive all-sky survey in the MIR including $W1$, $W2$, $W3$, and $W4$ 
centered at 3.4, 4.6, 12, and 22~\um, of which the angular resolution is 6.1, 6.4, 
6.5, and 12.0\arcsec, respectively. The 5$\sigma$ point source achieved by 
$WISE$ has sensitivities better than 0.08, 0.11, 1, and 6~mJy (corresponding to 
Vega magnitudes of 16.5, 15.5, 11.2, and 7.9) at these four bands, respectively, 
in unconfused regions on the ecliptic plane. We cross-match our sample with 
the $WISE$ All-Sky catalog using a radius of 6\arcsec \citep[e.g.,][]{wright10,
wenxq13}, which results in a subsample of 14,545 objects.

The $ROSAT$ All-Sky Survey (RASS; \citealp{boller16}) was the first to scan 
the entire celestial sphere in the 0.1--2.4 keV range with the Position Sensitive 
Proportional Counter (PSPC; \citealp{pfeffermann87}). The typical flux limit is 
typically  a few times $10^{-13}$~\flux\ and the positional uncertainties are 
typically 10--30 arcsec. We search for a possible RASS counterpart of our sample 
objects with a matching radius of 30\arcsec\ \citep[e.g.,][]{shensy06,
anderson07}. Given the relatively large uncertainty in the fluxes of the RASS 
objects, we take the pointed observational data (2RXP) into account as well 
whenever available. For each source detected in both the surveys, the data from 
the pointed observation are adopted. A total of 4221 objects have 
\rosat\,measurements.

The radio properties are derived from the VLA FIRST survey \citep{white97}, 
using the Very Large Array with a typical rms of 0.15~mJy, and a resolution of 
5\arcsec. About 90 objects are detected in each square degree with a detection 
threshold of 1~mJy, and $\sim35$\% of these are resolved on scales from 2\arcsec 
to 30\arcsec. We match our DR7 AGN catalog with the FIRST catalog using two 
different matching radii: one of 5\arcsec, for core-dominant radio sources, and 
the other is with a matching radius of 30\arcsec\ for lobe-dominant radio sources 
\citep{sheny11}. The radio detection fraction is $\sim$12\%, consistent with 
10\%--15\% of Seyfert galaxies and quasars in previous literature \citep[e.g.,][]{ivezic02}. 

\input{table3.tex}

The format of Part~III of the AGN catalog is as follows.
\begin{enumerate}
\item[1.] Identification number assigned in this paper.  
\item[2-11.] The SDSS PSF magnitudes in {\it ugriz} and their uncertainties.
\item[12-21.] The SDSS Petrosian magnitudes in {\it ugriz} and their 
uncertainties.
\item[22-25.] Magnitudes in NUV and FUV and their uncertainties given by 
GALEX.
\item[26-37.] Magnitudes in $J$, $H$, and $K_{\rm s}$, and their uncertainties 
given by 2MASS.
\item[38-45.] Magnitudes in $W1$, $W2$, $W3$, and $W4$, and their 
uncertainties given by $WISE$.
\item[46-47.] $ROSAT$ count rate (count~s$^{-1}$) and its uncertainty.
\item[48-51.] $F_{\rm peak}$ and $F_{\rm int}$ in units of mJy, and their 
uncertainties given by FIRST.  
$F_{\rm peak}$ and $F_{\rm int}$ are the peak and integrated flux densities at 
20~cm, which are estimated by fitting an elliptical Gaussian model to the source. 
Note that 0.25~mJy has been added to the peak flux density and the integrated 
flux density has been multiplied by ($1+0.25$/$F_{\rm peak}$),  in order to 
correct for the so-called CLEAN bias effect; for details see \citet{white97}. The 
uncertainty in $F_{\rm peak}$ is given by the rms noise at the source position, 
while the uncertainty in $F_{\rm int}$ can be considerably greater depending on 
the source size and morphology. The significance of detection for a source is 
($F_{\rm peak}$-0.25)/rms due to the CLEAN bias correction.
\end{enumerate}

\section{SUMMARY}
Large and homogeneous optical spectroscopic surveys such as the SDSS 
enable a census of active BHs located at the centers of galaxies over a wide 
range of physical parameters. In this work, we carry out a systematic and 
comprehensive analysis of the spectroscopic data from the tremendous galaxy 
and quasar database in the SDSS DR7, based on the matured spectral 
modeling procedures and algorithms that have been developed over the last 
decades \citep{dongxb05,dongxb08,dongxb12,zhouhy06,liuhy18}. Elaborate 
attention is paid to proper decomposition of the stellar continua and accurate 
deblending of the broad and narrow lines. The stellar templates are built
by applying the EL-ICA  algorithm to SSPs and emission lines are modeled and 
measured from the residual pure emission-line spectrum after subtracting the 
pseudo-continuum. In order to obtain proper broad-line measurements, a series 
of narrow-line models are used and the one with the minimum reduced \chisq\ is 
adopted as the best fit. Eventually, a large sample of 14,584 broad-line AGNs 
with $z<0.35$ is identified based on our well-defined broad-line criteria. We 
construct a catalog of all the sample objects in which various parameters are 
derived and tabulated, including the continuum and emission line parameters 
from spectral fits as well as derived BH mass and Eddington ratio. Moreover, 
photometric data in multiwavelength bands, and objects of special interest 
(spectra showing peculiar profiles) are flagged. 

This AGN catalog, homogeneously selected, along with the accurately measured 
spectral parameters, provides the most updated, largest AGN sample data, 
which will enable further comprehensive investigations of the properties of the 
AGN population in the low-redshift universe, such as the BH mass function 
(W.-J. Liu et al. 2020, in preparation), multiwavelength SED, and connection 
between SMBHs and host galaxies.
\acknowledgments
This work is supported by the National Natural Science Foundation of China 
(Grant No.\,11473035, No.\,11873083 and No.\,Y811011N05), the Strategic 
Pioneer Program on Space Science, Chinese Academy of Sciences (Grant 
No.XDA15052100) and the National Program on Key Research, and 
Development Project (Grant No.\,2016YFA0400804). 
W.L. acknowledges supports from the Natural Science Foundation of China 
grant (NSFC 11703079) and the ``Light of West China'' Program of Chinese 
Academy of Sciences (CAS) and the funding from Key Laboratory of Space 
Astronomy and Technology, National Astronomical Observatories, Chinese 
Academy of Sciences, Beijing 100012, China. T.W. acknowledges supports 
from the NSFC-CAS joint fund of astronomy U1431229. We are grateful to the 
anonymous referee for his/her constructive comments that improved the paper. 
This work is mainly based on the observations obtained by the SDSS, we 
acknowledge the entire SDSS team for providing the data that made this work 
possible. This publication also makes use of data products from GALEX, $WISE$, 
2MASS, $ROSAT$, and FIRST. GALEX is by the National Aeronautics and 
Space Administration (NASA). $WISE$ is a joint project of the University of 
California, Los Angeles, and the Jet Propulsion Laboratory/California Institute of 
Technology, funded by NASA. 2MASS is a joint project of the University of 
Massachusetts and the Infrared Processing and Analysis Center/California 
Institute of Technology, funded by the NASA and the National Science 
Foundation (NSF). We use of the $ROSAT$ Data Archive of the 
Max-Planck-Institut f{\"u}r extraterrestrische Physik (MPE) at Garching, Germany. FIRST is 
supported by National Radio Astronomy Observatory (NRAO).

\bibliographystyle{aasjournal}
\bibliography{liuheyang}
\clearpage

\end{document}

%% file: table1.tex
\begin{deluxetable*}{rccclcccc}
\tablecolumns{9} 
\tablecaption{The SDSS DR7 Broad-line AGN Sample (Part I)} %
\tablehead{ \colhead{ID}  & \colhead{Designation} & \colhead{R.A.} & \colhead{Decl.} & \colhead{$z$} &
\colhead{Plate} & \colhead{MJD} & \colhead{FiberID} & \colhead{$N_{\rm spec}$}
\\
\colhead{(1)} & \colhead{(2)} & \colhead{(3)} & \colhead{(4)} &
\colhead{(5)} & \colhead{(6)} & \colhead{(7)} & \colhead{(8)} &
\colhead{(9)}
}
\startdata
  1    & J000048.16$-$095404.0 &  0.20064719 &  $-$9.9011221  &   0.205  &  650  &   52143 &  494   &    1 \\
  2    & J000102.19$-$102326.9 &  0.2591277  &  $-$10.390799  &   0.294  &  650  &   52143 &  166   &    1 \\
  3    & J000111.15$-$100155.5 &  0.29645887 &  $-$10.032093  &   0.049 &  650  &   52143 &  174   &    1
\enddata
\tablecomments{
Col. (1): Identification number assigned in this paper.
Col. (2): Official SDSS designation in J2000.0.
Col. (3): Right ascension in decimal degrees (J2000.0).
Col. (4): Declination in decimal degrees (J2000.0).
Col. (5): Redshift measured by the SDSS pipeline.
Col. (6): Spectroscopic plate number.
Col. (7): Modified Julian date (MJD) of spectroscopic observation.
Col. (8): Spectroscopic fiber number.
Col. (9): Number of spectroscopic observations.
({\it This table is available in its entirety in a machine-readable form in the online
journal. A portion is shown here for guidance regarding its form and content.})
}
\label{tab1}
\end{deluxetable*}

%% file: table2.tex
\begin{deluxetable*}{clcl}
\tablecolumns{4} \tabletypesize{\scriptsize} \tablewidth{0pc}
\tablecaption{Catalog Format for the SDSS DR7 broad-line AGN sample (Part II)} %
\tablehead{ \colhead{Column} & \colhead{Name}  & \colhead{Format} 
& \colhead{Description} }
\startdata
	 1  & ID						& LONG		& Identification number assigned in this paper. \\
	 2  & CENTROID\_BHA				& DOUBLE 	& Centroid wavelength of broad \ha\ (\AA) \\
	 3	& CENTROID\_BHA\_ERR		& DOUBLE 	& Uncertainty in Centroid wavelength of broad \ha\ (\AA) \\
     4  & LBHA						& DOUBLE    & Line luminosity of broad \ha\ [$\log (L_{{\rm H\alpha,broad}}/\rm erg~s^{-1})$]\\
 	 5  & LBHA\_ERR					& DOUBLE   	& Uncertainty in $\log L_{{\rm H\alpha,broad}}$\\
 	 6  & FWHM\_BHA					& DOUBLE    & FWHM of broad \ha\ (\kms)\\
 	 7  & FWHM\_BHA\_ERR			& DOUBLE    & Uncertainty in the ${\rm FWHM_{H\alpha,broad}}$ \\
 	 8  & EW\_BHA					& DOUBLE    & Restframe equivalent width of broad \ha\ (\AA)\\
 	 9  & EW\_BHA\_ERR				& DOUBLE    & Uncertainty in ${\rm EW_{H\alpha,broad}}$\\
	10  & CENTROID\_NHA				& DOUBLE 	& Centroid wavelength of narrow \ha\ (\AA) \\
	11  & CENTROID\_NHA\_ERR		& DOUBLE 	& Uncertainty in Centroid wavelength of narrow \ha\ (\AA) \\
 	12  & LNHA						& DOUBLE    & Line luminosity of narrow \ha\ [$\log (L_{{\rm H\alpha,narrow}}/\rm erg~s^{-1})$] \\
 	13  & LNHA\_ERR					& DOUBLE   	& Uncertainty in $\log L_{{\rm H\alpha,narrow}}$\\
 	14  & FWHM\_NHA					& DOUBLE    & FWHM of narrow \ha\ (\kms)\\
 	15  & FWHM\_NHA\_ERR			& DOUBLE    & Uncertainty in the ${\rm FWHM_{H\alpha,narrow}}$ \\
 	16  & EW\_NHA 					& DOUBLE    & Restframe equivalent width of narrow \ha\ (\AA)\\
 	17  & EW\_NHA\_ERR				& DOUBLE   	& Uncertainty in ${\rm EW_{H\alpha,narrow}}$\\
 	18  & LNII6583					& DOUBLE   	& Line luminosity of $\rm \nii\lambda$6583 [$\log (L_{{\rm \nii\lambda6583}}/\rm erg~s^{-1})$]\\
 	19  & LNII6583\_ERR				& DOUBLE   	& Uncertainty in $\log L_{{\rm \nii\lambda6583}}$ \\
 	20  & EW\_NII6583				& DOUBLE   	& Restframe equivalent width of $\rm \nii\lambda$6583 (\AA) \\
 	21  & EW\_NII6583\_ERR 			& DOUBLE   	& Uncertainty in ${\rm EW_{\nii\lambda6583}}$ \\
 	22  & LSII6716					& DOUBLE   	& Line luminosity of $\rm \sii\lambda$6716 [$\log (L_{{\rm \sii\lambda6716}}/\rm erg~s^{-1})$]\\
 	23  & LSII6716\_ERR				& DOUBLE   	& Uncertainty in $\log L_{\rm \sii\lambda6716}$\\
 	24  & EW\_SII6716				& DOUBLE   	& Restframe equivalent width of $\rm \sii\lambda$6716 (\AA)\\
 	25  & EW\_SII6716\_ERR 			& DOUBLE   	& Uncertainty in ${\rm EW_{\sii\lambda6716}}$\\
 	26  & LSII6731					& DOUBLE   	& Line luminosity of $\rm \sii\lambda$6731 [$\log (L_{{\rm \sii\lambda6731}}/\rm erg~s^{-1})$]\\
 	27  & LSII6731\_ERR				& DOUBLE   	& Uncertainty in $\log L_{\rm \sii\lambda6731}$\\
 	28  & EW\_SII6731				& DOUBLE   	& Restframe equivalent width of $\rm \sii\lambda$6731 (\AA)\\
 	29  & EW\_SII6731\_ERR 			& DOUBLE   	& Uncertainty in ${\rm EW_{\sii\lambda6731}}$\\
 	30  & LOI6300					& DOUBLE   	& Line luminosity of $\rm \oi\lambda$6300 [$\log (L_{{\rm \oi\lambda6300}}/\rm erg~s^{-1})$]\\
 	31  & LOI6300\_ERR				& DOUBLE   	& Uncertainty in $\log L_{\rm \oi\lambda6300}$\\
 	32  & EW\_OI6300				& DOUBLE   	& Restframe equivalent width of $\rm \oi\lambda$6300 (\AA)\\
 	33  & EW\_OI6300\_ERR 			& DOUBLE   	& Uncertainty in ${\rm EW_{\oi\lambda6300}}$\\
	34  & Centroid\_BHB				& DOUBLE 	& Centroid wavelength of broad \hb\ (\AA) \\
	35  & Centroid\_BHB\_ERR		& DOUBLE 	& Uncertainty in Centroid wavelength of broad \hb\ (\AA) \\
    36  & LBHB						& DOUBLE    & Line luminosity of broad \hb\ [$\log (L_{{\rm H\beta,broad}}/\rm erg~s^{-1})$]\\
 	37  & LBHB\_ERR					& DOUBLE   	& Uncertainty in $\log L_{{\rm \hb,broad}}$\\
 	38  & FWHM\_BHB					& DOUBLE    & FWHM of broad \hb\ (\kms)\\
 	39  & FWHM\_BHB\_ERR			& DOUBLE    & Uncertainty in the ${\rm FWHM_{\hb,broad}}$ \\
 	40  & EW\_BHB					& DOUBLE    & Restframe equivalent width of broad \hb\ (\AA)\\
 	41  & EW\_BHB\_ERR				& DOUBLE    & Uncertainty in ${\rm EW_{\hb,broad}}$\\
	42  & CENTROID\_NHB				& DOUBLE 	& Centroid wavelength of narrow \hb\ (\AA) \\
 	43  & LNHB						& DOUBLE    & Line luminosity of narrow \hb\ [$\log (L_{{\rm \hb,narrow}}/\rm erg~s^{-1})$]\\
 	44  & LNHB\_ERR					& DOUBLE   	& Uncertainty in $\log L_{{\rm \hb,narrow}}$\\
 	45  & EW\_NHB 					& DOUBLE    & Restframe equivalent width of narrow \hb\ (\AA)\\
 	46  & EW\_NHB\_ERR				& DOUBLE   	& Uncertainty in ${\rm EW_{\hb,narrow}}$\\
	47  & CENTROID\_OIII5007		& DOUBLE 	& Centroid wavelength of $\rm \oiii\lambda$5007 (\AA) \\
	48  & CENTROID\_OIII5007\_ERR	& DOUBLE 	& Uncertainty in Centroid wavelength of $\rm \oiii\lambda$5007 (\AA) \\
 	49  & LOIII5007					& DOUBLE    & Line luminosity of $\rm \oiii\lambda$5007 [$\log (L_{{\rm \oiii5007}}/\rm erg~s^{-1})$]\\
 	50  & LOIII5007\_ERR			& DOUBLE   	& Uncertainty in $\log L_{\rm \oiii5007}$ \\
 	51  & FWHM\_OIII5007			& DOUBLE    & FWHM of $\rm \oiii\lambda$5007 (\kms)\\
 	52  & FWHM\_OIII5007\_ERR		& DOUBLE    & Uncertainty in the ${\rm FWHM_{\oiii\lambda5007}}$  \\
 	53  & EW\_OIII5007 				& DOUBLE   	& Restframe equivalent width of $\rm \oiii\lambda$5007 (\AA) \\
 	54  & EW\_OIII5007\_ERR 	  	& DOUBLE  	& Uncertainty in ${\rm EW_{\oiii\lambda5007}}$ \\
 	55  & LFeII4570					& DOUBLE    & Line luminosity of $\rm \feii\lambda$4570 [$\log (L_{{\rm \feii4570}}/\rm erg~s^{-1})$]\\
 	56  & LFeII4570\_ERR			& DOUBLE   	& Uncertainty in $\log L_{\rm \feii4570}$ \\
 	57  & EW\_FeII4570 				& DOUBLE   	& Restframe equivalent width of $\rm \feii\lambda$4570 (\AA) \\
 	58  & EW\_FeII4570\_ERR 	  	& DOUBLE  	& Uncertainty in ${\rm EW_{\feii\lambda4570}}$ \\
	59  & L5100						& DOUBLE	& Monochromatic luminosity at 5100~\AA\ in log-scale \\
	60  & L5100\_ERR				& DOUBLE	& Uncertainty in monochromatic luminosity at 5100~\AA\ in log-scale \\
	61  & MultiPeak     	    	& Long 		& 0 = no peculiar profile; 1 = $\rm \oiii\lambda$5007 with multiple peak profile; \\
	  	&							&			& 2 = broad Balmer line with multiple peak profile \\
 	62  & MBH\_BHB   				& DOUBLE  	& Virial black hole mass in log-scale based on \hb\ \citep{ho15} \\
 	63  & MBH\_BHA   				& DOUBLE  	& Virial black hole mass in log-scale based on \ha\ (derived from \citealp{ho15} and \citealp{greene05b}) \\
 	64  & MBH   					& DOUBLE  	& The adopted fiducial virial black hole mass in log-scale \\
 	65  & LAMBDA\_EDD   			& DOUBLE  	& Eddington ratio based on the fiducial virial BH mass ($\log$ \lbol/\ledd) \\
\enddata
\label{tab2}
\tablecomments{This table is available in its entirety in a machine-readable form in the online journal.}
\end{deluxetable*}

%% file: table3.tex
\begin{deluxetable*}{clcl}
\tablecolumns{4} \tabletypesize{\scriptsize} \tablewidth{0pc}
\tablecaption{Catalog Format for the SDSS DR7 broad-line AGN sample (Part III)} %
\tablehead{ \colhead{Column} & \colhead{Name}  & \colhead{Format} 
& \colhead{Description} }
\startdata
	 1  & ID					& LONG		& Identification number assigned in this paper. \\
	 2  & PSFMAG\_u				& DOUBLE	& PSF magnitude in $u$ band, uncorrected for Galactic extinction \\
	 3  & PSFMAGERR\_u			& DOUBLE	& Uncertainty in $u$-band PSF magnitude \\
	 4  & PSFMAG\_g				& DOUBLE	& PSF magnitude in $g$ band, uncorrected for Galactic extinction \\
	 5  & PSFMAGERR\_g			& DOUBLE	& Uncertainty in $g$-band PSF magnitude \\
	 6  & PSFMAG\_r				& DOUBLE	& PSF magnitude in $r$ band, uncorrected for Galactic extinction \\
	 7  & PSFMAGERR\_r			& DOUBLE	& Uncertainty in $r$-band PSF magnitude \\
	 8  & PSFMAG\_i				& DOUBLE	& PSF magnitude in $i$ band, uncorrected for Galactic extinction \\
	 9  & PSFMAGERR\_i			& DOUBLE	& Uncertainty in $i$-band PSF magnitude \\
	10  & PSFMAG\_z				& DOUBLE	& PSF magnitude in $z$ band, uncorrected for Galactic extinction \\
	11  & PSFMAGERR\_z			& DOUBLE	& Uncertainty in $z$-band PSF magnitude \\
	12  & PETROMAG\_u			& DOUBLE	& Petrosian magnitude in $u$ band, uncorrected for Galactic extinction \\
	13  & PETROMAGERR\_u		& DOUBLE	& Uncertainty in $u$-band Petrosian magnitude \\
	14  & PETROMAG\_g			& DOUBLE	& Petrosian magnitude in $g$ band, uncorrected for Galactic extinction \\
	15  & PETROMAGERR\_g		& DOUBLE	& Uncertainty in $g$-band Petrosian magnitude \\
	16  & PETROMAG\_r			& DOUBLE	& Petrosian magnitude in $r$ band, uncorrected for Galactic extinction \\
	17  & PETROMAGERR\_r		& DOUBLE	& Uncertainty in $r$-band Petrosian magnitude \\
	18  & PETROMAG\_i			& DOUBLE	& Petrosian magnitude in $i$ band, uncorrected for Galactic extinction \\
	19  & PETROMAGERR\_i		& DOUBLE	& Uncertainty in $i$-band Petrosian magnitude \\
	20  & PETROMAG\_z			& DOUBLE	& Petrosian magnitude in $z$ band, uncorrected for Galactic extinction \\
	21  & PETROMAGERR\_z		& DOUBLE	& Uncertainty in $z$-band Petrosian magnitude \\
	22  & MAG\_FUV				& DOUBLE    & Magnitude in GALEX FUV ($\lambda_{\rm eff}=1528$~\AA), uncorrected for Galactic extinction \\
	23  & MAGERR\_FUV			& DOUBLE    & Uncertainty in GALEX FUV magnitude \\
	24  & MAG\_NUV				& DOUBLE    & Magnitude in GALEX NUV ($\lambda_{\rm eff}=2271$~\AA), uncorrected for Galactic extinction \\
	25  & MAGERR\_NUV			& DOUBLE    & Uncertainty in GALEX NUV magnitude \\
	26  & MAG\_J				& DOUBLE    & Magnitude in 2MASS $J$ band (1.26~\um) derived from PSC\\
	27  & MAGERR\_J				& DOUBLE    & Uncertainty in $J$-band magnitude derived from PSC\\
	28  & MAG\_H				& DOUBLE    & Magnitude in 2MASS $H$ band (1.65~\um) derived from PSC\\
	29  & MAGERR\_H				& DOUBLE    & Uncertainty in $H$-band magnitude derived from PSC\\
	30  & MAG\_Ks				& DOUBLE    & Magnitude in 2MASS $K_s$ band (2.16~\um) derived from PSC\\
	31  & MAGERR\_Ks			& DOUBLE    & Uncertainty in $K_s$-band magnitude derived from PSC\\
	32  & MAG\_J\_EXT			& DOUBLE    & Magnitude in 2MASS $J$ band (1.26~\um) derived from XSC\\
	33  & MAGERR\_J\_EXT		& DOUBLE    & Uncertainty in $J$-band magnitude derived from XSC\\
	34  & MAG\_H\_EXT			& DOUBLE    & Magnitude in 2MASS $H$ band (1.65~\um) derived from XSC\\
	35  & MAGERR\_H\_EXT		& DOUBLE    & Uncertainty in $H$-band magnitude derived from XSC\\
	36  & MAG\_Ks\_EXT			& DOUBLE    & Magnitude in 2MASS $K_s$ band (2.16~\um) derived from XSC\\
	37  & MAGERR\_Ks\_EXT		& DOUBLE    & Uncertainty in $K_s$-band magnitude derived from XSC\\
	38  & MAG\_W1				& DOUBLE    & Magnitude in $WISE$ $W1$ band (3.4~\um) \\
	39  & MAGERR\_W1			& DOUBLE    & Uncertainty in $WISE$ $W1$-band magnitude \\
	40  & MAG\_W2				& DOUBLE    & Magnitude in $WISE$ $W2$ band (4.6~\um) \\
	41  & MAGERR\_W2			& DOUBLE    & Uncertainty in $WISE$ $W2$-band magnitude \\
	42  & MAG\_W3				& DOUBLE    & Magnitude in $WISE$ $W3$ band (12~\um) \\
	43  & MAGERR\_W3			& DOUBLE    & Uncertainty in $WISE$ $W3$-band magnitude \\
	44  & MAG\_W4				& DOUBLE    & Magnitude in $WISE$ $W4$ band (22~\um) \\
	45  & MAGERR\_W4			& DOUBLE    & Uncertainty in $WISE$ $W4$-band magnitude \\
	46  & COUNT					& DOUBLE    & \rosat\ count rate in 0.1--2.4~keV (count s$^{-1}$) \\
	47  & COUNTERR			& DOUBLE    & Uncertainty in the \rosat\ X-ray count rate \\
	48	& FLAG\_FIRST			& DOUBLE	& FIRST flag ($-1=$~not in FIRST footprint; 0~$=$~FIRST undetected; 1~$=$~core-dominant; 2~$=$~lobe-dominant) \\
	49  & Fpeak					& DOUBLE    & Peak flux density at 20~cm from FIRST (mJy) \\
	50  & Fint					& DOUBLE    & Integrated flux density at 20~cm from FIRST (mJy) \\
	51  & RMS					& DOUBLE    & Local noise estimate at the source position measured by FIRST (mJy) 
\enddata
\label{tab3}
\tablecomments{This table is available in its entirety in a machine-readable form in the online journal.}
\end{deluxetable*}